\RequirePackage{fix-cm}
\documentclass{svjour3}                     
\usepackage{graphicx}
\usepackage{tabularx}
\usepackage{natbib,twoopt}
\bibpunct{(}{)}{;}{a}{}{,}             
\usepackage{amssymb}
\usepackage{gensymb}
\begin{document}

\title{A better characterization of the chemical composition of exoplanets atmospheres with ARIEL
}


\author{Olivia Venot \and Benjamin Drummond \and Yamila Miguel \and Ingo P. Waldmann \and Enzo Pascale \and Tiziano Zingales}


\institute{O. Venot \at
              Laboratoire Interuniversitaire des Syst\`{e}mes Atmosph\'{e}riques, UMR CNRS 7583, Universit\'{e} Paris Est Cr\'eteil (UPEC) et Universit\'e Paris Diderot (UPD), Institut Pierre Simon Laplace (IPSL), Cr\'{e}teil, France \\
              \email{olivia.venot@lisa.u-pec.fr}           
            \and
            B. Drummond \at
            Astrophysics Group, University of Exeter, EX4 4QL, Exeter, UK
            \and
            Y. Miguel \at
            Laboratoire Lagrange, UMR 7293, Universit\'e de Nice-Sophia Antipolis, CNRS, Observatoire de la C\^ote d'Azur, Blvd de l'Observatoire, CS 34229, 06304 Nice cedex 4, France, and \\ 
            Leiden Observatory, University of Leiden, Niels Bohrweg 2, 2333CA Leiden, The Netherlands
            \and I. P. Waldmann \at University College London, Department of Physics and Astronomy, Gower Street, London WC1E 6BT, UK
            \and E. Pascale \at Dipartimiento di Fisica, La Sapienza Universit\`a di Roma, Piazzale Aldo Moro 2, 00185 Roma, Italy, and\\
            Cardiff University, School of Physics and Astronomy, 5 The Parade, Cardiff, CF24 3AA, UK
            \and T. Zingales \at University College London, Department of Physics and Astronomy, Gower Street, London WC1E 6BT, UK, and \\
            INAF--Osservatorio Astronomico di Palermo, Piazza del Parlamento 1, I-90134 Palermo, Italy
}

\date{Received: date / Accepted: date}

\maketitle

\begin{abstract}
Since the discovery of the first extrasolar planet more than twenty years ago, nearly four thousand planets orbiting stars other than the Sun\footnote{http://exoplanet.eu/} have been discovered. Current observational instruments (on board the Hubble Space Telescope, Spitzer, and on ground-based facilities) have allowed the scientific community to obtain important information on the physical and chemical properties of these planets. However, for a more in-depth characterisation of these worlds, more powerful telescopes are needed. Thanks to the high sensitivity of their instruments, the next generation of space observatories (e.g. JWST, ARIEL) will provide observations of unprecedented quality, allowing us to extract far more information than what was previously possible. Such high quality observations will provide constraints on theoretical models of exoplanet atmospheres and lead to a greater understanding of their physics and chemistry. Important modelling efforts have been carried out during the past few years, showing that numerous parameters and processes (such as the elemental abundances, temperature, mixing, etc.) are likely to affect the atmospheric composition of exoplanets and subsequently the observable spectra. In this manuscript, we review the different parameters that can influence the molecular composition of exoplanet atmospheres. We show that the high-precision of ARIEL observations will improve our view and characterisation of exoplanet atmospheres. We also consider future developments that are necessary to improve atmospheric models, driven by the need to interpret the available observations.

\keywords{Atmospheres \and Exoplanets \and Composition \and Modelling \and Laboratory measurements \and Observations}
\end{abstract}

\section{Introduction}
\label{intro}
The number of known exoplanets has exploded during the past few years. It is now clear that there exists a huge diversity of worlds in terms of radius, mass, temperature, orbital eccentricity, etc. This wealth of new information has implications for planetary formation scenarios, originally formulated in the time when only the Solar system planets were known. The chemical composition of the atmosphere is the key to understanding planetary formation. In particular, the elemental composition is expected to depend on the environment in which the planet formed, and through which it evolved. Thanks to spectroscopic observations performed during planetary transits, the atmosphere of planets can be probed, giving access to their chemical composition. 

To a fundamental level, the chemical composition of the atmosphere is determined by 1) the elemental abundances the planet formed with and 2) the temperature of the atmosphere, which is dependent on irradiation and internal heating. In addition, physical processes such as mixing, photolysis, and atmospheric escape, occur in different pressure regions of exoplanet atmospheres (see Fig. 1 of \citealt{mad2016}), and can influence the atmospheric composition, with subsequent effects on the observable spectra.  Infrared (IR) observations probe the middle atmosphere ($\sim 0.1 - 10^{-5}$ bar), giving information on the abundances of neutral species, whereas ultraviolet (UV) observations are used to study the upper atmosphere, where atomics species and ions are highly present. Models that couple chemical kinetics to the out of equilibrium processes mentioned above are used to predict the chemical abundances and to understand the spectroscopic observations. 

Several models have been developed up to now \citep[e.g.][]{liang2003,liang2004,koskinen2007,zahnle2009arXiv,zahnle2009,line2010,moses2011,hu2012,kopparapu2012,Venot_2012,agu2012,lavvas2014,miguel2014,drummond2016,rimmer2016,tsai2017} which differ in the complexity, size (number of species and reactions) and type (neutral/ionic) of chemistry, the temperature range over which the chemical scheme is valid, and the atmospheric layers that can be studied with (e.g. middle atmosphere, upper atmosphere). In this manuscript, we focus on the chemistry of neutral species. Depending on their abundance, these latter will have spectral signatures in the future observations of ARIEL (Atmospheric Remote-sensing Exoplanet Large-survey).

Among the different categories of exoplanets, the warm gaseous giant planets are the most interesting. Firstly, they provide the highest quality observations that are necessary for model--observation comparisons. Secondly, the abundances of the major molecules (H$_2$O, CO, NH$_3$, etc.) constrained by observations might be a direct reflection of the elemental abundances of the planetary atmosphere, unlike giant planets of the Solar System. Indeed, the abundance of oxygen in the atmosphere of Jupiter or Uranus is highly uncertain because the condensation of water removes oxygen species from the observable part of the atmosphere \citep{cavalie2017}. However, in the much hotter atmospheres of short-period exoplanets, condensation is limited to high temperature condensate species, such as TiO or VO, which are not the main carrier of elemental oxygen. Finally, the atmospheres of gas giants may be similar to the elemental composition with which they formed; unlike smaller planets which may be affected by delivery from meteorites/comets or out-gassing from the rocky interior. We precise that determining the enrichment of the planetary atmosphere relative to the elemental composition of the star requires in addition good observations of the host star.

The goal of this manuscript is to consider some important model parameters that effect the atmospheric composition and, most importantly, to test whether these effects are significant enough to be detected with ARIEL. We consider several previously published studies and calculate new synthetic transmission and emission spectra, taking into account the technical characteristics of the instruments on board ARIEL.

In Section \ref{sec:1Dmodel}, we present the tools that have been used in this paper: a series of one-dimensional kinetic models, a forward radiative-transfer code and the ARIEL noise model. In Section \ref{sec:eq-neq} we describe the difference between equilibrium and disequilibrium chemistry regimes. In Sections \ref{section:model_parameters}, \ref{sec:star_param} and \ref{sec:prog_mod} we consider how various planetary, orbital and stellar parameters effect the synthetic spectra, specifically in the context of ARIEL. Finally, we look forward to potential advances in understanding that will be achieved with the ESA ARIEL space mission.

\section{Models used in this work}\label{sec:1Dmodel}

\subsection{One-dimensional chemical kinetics models}

The results presented in this study are obtained using a number of different chemical models. These one-dimensional, time-dependent models include chemical kinetics as well as processes that drive the chemistry away from local chemical equilibrium (vertical mixing and photochemistry). These models are fully described in: \citet{Venot_2012}, \citet{miguel2014}, \citet{agu2014} and \citet{drummond2016}.

Generally, these models represent the atmosphere as a vertical column with a given thermal profile. This profile is typically divided into several layers with a thickness equal to a constant fraction of the pressure scale height; except for \citet{drummond2016} where the layers are determined directly by hydrostatic balance. To determine the chemical composition of the column, the continuity equation (Eq.~\ref{eq:continuity}) is solved for the steady-state of each species in each model level, 

\begin{equation}\label{eq:continuity}
\frac{\partial n_i}{\partial t} = P_i - n_iL_i - div({\Phi_i}\overrightarrow{e_z})
\end{equation}
where $n_i$ the number density ($\mathrm{cm^{-3}}$), $P_i$ the production rate ($\mathrm{cm^{-3}s^{-1}}$), $L_i$ the loss rate ($\mathrm{s^{-1}}$),  and $\Phi_i$ the vertical flux ($\mathrm{cm^{-2}s^{-1}}$), respectively, of the species $i$. The vertical flux is parameterised by the vertical diffusion equation,

\begin{equation}
\Phi_i = -n_iD_i \left[ \frac{1}{n_i}\frac{\partial n_i}{\partial z}+\frac{1}{H_i}+\frac{1}{T}\frac{dT}{dz}\right]-n_iK\left[\frac{1}{y_i}\frac{\partial y_i}{\partial z}\right],
\end{equation}
where $K$ is the eddy diffusion coefficient ($\mathrm{cm^2s^{-1}}$), $D_i$ is the molecular diffusion coefficient ($\mathrm{cm^2s^{-1}}$), $H_i$ the scale height, and $y_i$ the mixing ratio of the species $i$.\\
At both upper and lower boundaries, a zero flux for each species is usually imposed.

One of the main ingredients of a kinetics model is the chemical network which, in essence, is a list of chemical reactions and associated rate constants. \citet{Venot_2012} implemented a chemical network entirely new in planetology, specifically adapted to the extreme conditions of warm exoplanet atmospheres. This scheme has been developed in close collaboration with experts in the combustion industry. Its strength comes from its global experimental validation that has been performed across a large range of pressures (0.01--100 bar) and temperatures (300--2500 K). The network describes the kinetics of species made of H, C, O, and N, including hydrocarbons with up to two carbon atoms. The 105 compounds of this scheme are linked by $\sim$2000 reactions. All the results presented in this paper have been found using this chemical scheme (i.e. \citealt{Venot_2014,Venot_2015,Venot_2016,Venot_2018,agu2012,agu2014,Agundez_2014,drummond2016}) except results of \cite{miguel2014} and \cite{miguel2015} that have been found using a smaller chemical scheme of 19 species and 179 reactions.

In order to study atmospheres rich in carbon, where hydrocarbons can be abundant and thus have a non-negligible influence on the chemical composition, \citet{Venot_2015} developed an extended version of the chemical network, able to describe kinetics of species containing up to six carbon atoms. This large scheme contains 240 species and $\sim$4000 reactions. Both of these schemes are available on the KIDA Database\footnote{http://kida.obs.u-bordeaux1.fr/}\citep{wakelam2012}.

\subsection{Radiative transfer model}
\label{sec:rt}

In this study we calculate a series a transmission and emission spectra using the radiative transfer forward model in the TauREx retrieval framework \citep{waldmann2015tau,waldmann2015rex}, using thermal and composition profiles from previously published works. The infrared absorption cross-sections of the absorbing species were computed using the linelists from ExoMol \citep{tennyson2012}, where available, and otherwise taken from HITEMP \citep{rothman2010hitemp} and HITRAN \citep{rothman2009,rothman2013}. Cross sections and subsequent forward models were computed at a resolution of R~=~7,000 (constant in wavelength). We assume 300 plane-parallel atmospheric layers and cloud-free atmospheres. To simulate the observations that ARIEL will provide, we bin the high-resolution forward model to the wavelength and resolution grid of ARIEL for the specified signal-to-noise of the observation (see Section \ref{sec:noise}).

\subsection{ARIEL noise model}\label{sec:noise}
The signal-to-noise ratio (SNR) has been evaluated using a radiometric model which implements a detailed description of the ARIEL instrument design described in details in the ARIEL Assessment Study Report (Yellow Book)\footnote{http://sci.esa.int/cosmic-vision/59109-ariel-assessment-study-report-yellow-book/}.\\
The noise model implemented provides an accurate estimate of the experimental uncertainties of the planet spectra ARIEL will measure. All major sources of photometric uncertainty are accounted for. Among these are: photodetector non uniform response; intra-pixel variations; pointing stability; photon shot noise; detector read noise. Models implemented are cross validated using the end-to-end time-domain transit-spectroscopy simulator ExoSim \citep{sarkar2016}, and using the radiometric model developed by ESA \citep{Puig2015}. Details of the cross-validation of these simulation tools are found in the ARIEL Yellow Book accompanying documentation.\\
In Table \ref{tab:char_ARIEL} we show the spectral coverage of ARIEL and the nominal resolution of each band.  In Table \ref{tab:SNR} we present the SNR per transit of the three planets studied in this paper.
\begin{table}[!h]
\caption{ARIEL spectral coverage and nominal resolution. The three first channels are photometric bands. The three last ones are covered by ARIEL spectrograph.} \label{tab:char_ARIEL}
\centering
\begin{tabular}{c|cc}
\hline \hline
Spectral range  & Wavelength range ($\mu$m)  & Resolution \\
\hline
1 & 0.5 -- 0.55 & Integrated band \\
\hline
2 & 0.8 -- 1.0 & Integrated band \\
\hline
3 & 1.05 -- 1.2 & Integrated band \\
\hline
\hline
4  & 1.25 -- 1.95 & 10 \\
\hline
5 & 1.95 -- 3.90 & 100 \\
\hline
6 & 3.90 -- 7.80 & 30 \\
 \hline
\end{tabular}
\end{table}
\begin{table}[!h]
\caption{SNR per transit for the planets of this study.} \label{tab:SNR}
\centering
\begin{tabular}{c|ccc}
\hline \hline
Spectral range  & HD 189733b  & GJ 436b & GJ 3470b \\
\hline
1  & 21.79 & 4.19 & 2.35\\
\hline
2  & 48.96 & 19.77 & 10.07\\
\hline
3  & 39.19 & 18.13 & 9.01\\
\hline
\hline
4  & 34.99 & 21.19 & 10.10\\
\hline
5 & 7.34 & 4.89 & 2.33 \\
\hline
6 & 6.87 & 5.02 & 2.30 \\
 \hline
\end{tabular}
\end{table}

\section{The transition between chemical equilibrium and disequilibrium}\label{sec:eq-neq}

Intense irradiation of short--period exoplanets leads to very high atmospheric temperatures of typically $T>1000$\,K. This might initially suggest that the chemical composition of these atmospheres could be described by thermochemical equilibrium, as such temperatures lead to fast chemical kinetics. The first models used to study exoplanet atmospheres assumed chemical equilibrium \citep[e.g.][]{burrows1999chemical, seager2000theoretical, sharp2007atomic, barman2007, burrows2007theoretical, burrows2008theoretical}. However, it was quickly realised that such assumptions were not favoured by observations. 

Physical processes, such as mixing and photodissociations, can influence the chemical composition \citep[e.g.][]{zahnle2009,line2010,moses2011,Venot_2012,miguel2014,drummond2016}. 
In the presence of vertical mixing, a competition exists between the chemical and mixing (or dynamical) processes. If the timescale of the chemistry is faster than that of the mixing, then the atmosphere will achieve chemical equilibrium. On the other hand, if the mixing timescale is faster than the chemical timescale then the atmosphere will be driven away from chemical equilibrium. The pressure level at which these two timescales are equal is usually called the quenching point \citep[i.e.][]{prinn1977}, and the equilibrium abundances at the quench determine the ``quenched abundances'' for all lower pressures.

It is obvious that depending on the balance of the dynamical and chemical timescales, the composition of the atmosphere can be altered. If processes like mixing and photochemistry lead to significant changes in the abundances of absorbing chemical species, they may lead to signatures in the observable spectra. In this case, photochemical models are required to successfully interpret the spectra and to constrain the chemical composition.

\begin{figure}[h!]
\begin{center}
\includegraphics[width=0.9\columnwidth]{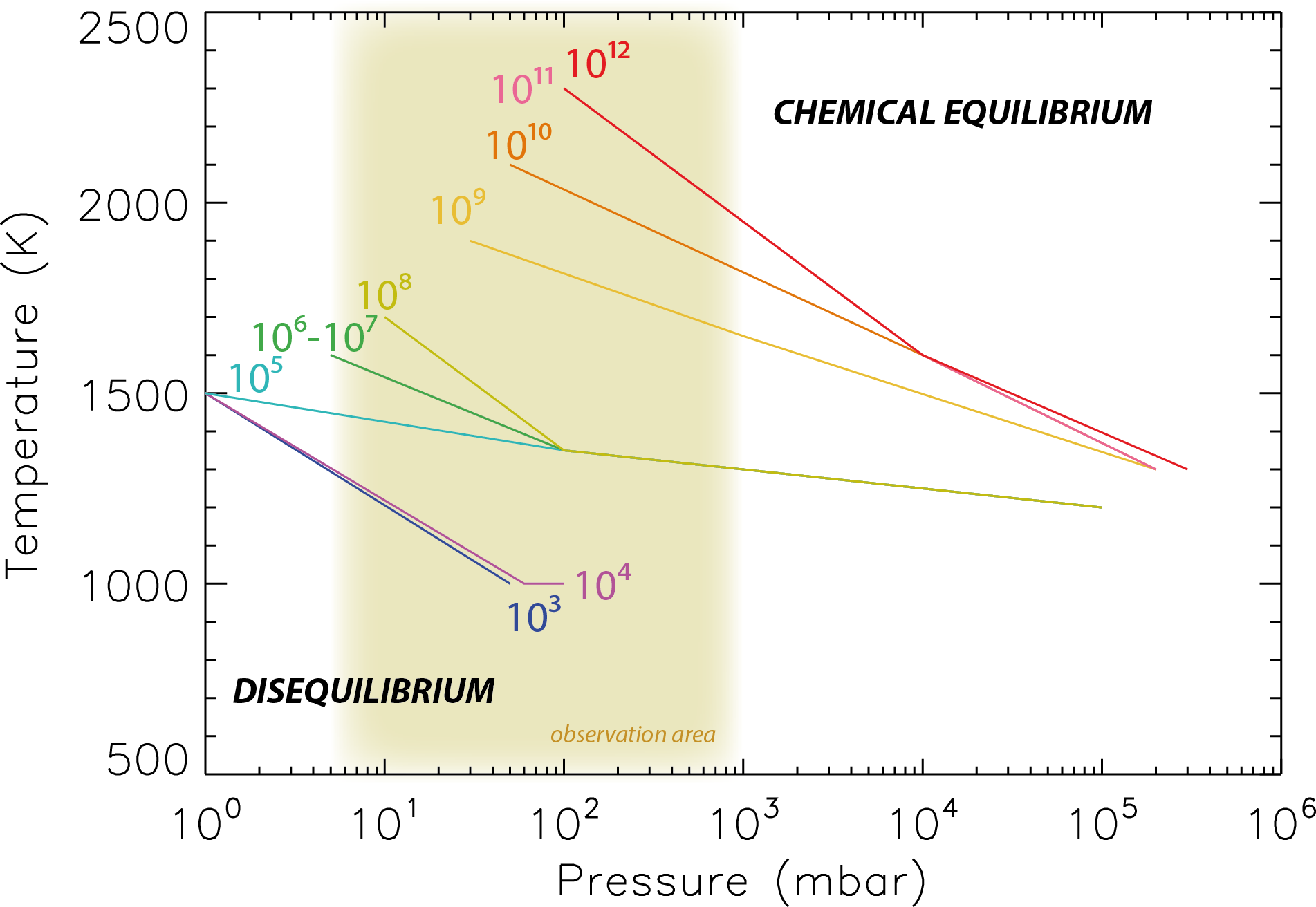}
\caption{{\label{eq/diseq}
Pressure-temperature chemical equilibrium/disequilibrium lines calculated for different vertical mixing strengths. The value of the eddy diffusion coefficient (in cm$^2$s$^{-1}$) are labeled on the figure and are represented by different colors. For instance, an atmospheric layer at 1500~K and 10$^2$~mbar is at chemical equilibrium if the vertical mixing is lower than 10$^9$~cm$^2$s$^{-1}$.%
}}
\end{center}
\end{figure}

We determined in which cases the atmosphere is likely to be in a state of chemical equilibrium, for a series of different thermal profiles published in \cite{Venot_2015}. These thermal profiles have been computed with the analytical model of \cite{parmentier2014a}. The three thermal profiles have each an isothermal part of 500, 1000, or 1500 K in the upper atmosphere, which correspond to irradiation temperatures of 784, 1522, or 2303~K, respectively (see \citealt{Venot_2015} and \citealt{parmentier2014a} for more details).\\ For each PT profile, we compared the steady-state abundances of the main species (i.e. H$_2$, H, H$_2$O, CH$_4$, CO, CO$_2$, N$_2$, NH$_3$, HCN, CH$_3$, and OH) with their thermochemical equilibrium abundances and determined their quenching levels. The deeper quench level defines the equilibrium/disequilibrium limit. Above this level, at least one species is not at thermochemical equilibrium. We reported the pressure and temperature of this quench point for each PT profile on Fig.~\ref{eq/diseq}. We also studied how the equilibrium/disequilibrium limit varies with vertical mixing strength, parameterised by the eddy diffusion coefficient $K_{zz}$, which typically varies between $10^3<K_{zz}<10^{12}$ cm$^2$s$^{-1}$.

Figure \ref{eq/diseq} shows the transition between chemical equilibrium and disequilibrium as a function of pressure and temperature, for different values of $K_{zz}$. Given that the pressure range probed by observations is approximately from 1 to 10$^3$ mbar, exoplanets with a temperature higher than 1500 K (in this pressure region) are more likely to be in chemical equilibrium, which, as we said previously, correspond to planets with an irradiation temperature of about 2300 K. For cooler planets disequilibrium is important and the use of a chemical kinetics model (accounting for vertical mixing) is necessary to determine the chemical composition of the atmosphere.

This evaluation is approximate because the equilibrium/disequilibrium transition depends on the value of $K_{zz}$, which is very uncertain. Nevertheless, eddy diffusion coefficients of about 10$^8$ cm$^2$s$^{-1}$ are commonly accepted in the community of exoplanets, based on calculations made with 3D models and parametrised to 1D calculations \citep{parmentier2013}. The transition location depends also on the shape of the thermal profile.
Two different carbon-to-oxygen ratios (C/O) have been tested (C/O solar and C/O=1.1), but this parameter has a negligible effect on the location of the equilibrium/disequilibrium transition.

\section{Influence of planetary and orbital parameters}
\label{section:model_parameters}

\subsection{Vertical mixing, metallicity, temperature}\label{sec:details_effect}

For warm planets whose atmospheres are unlikely to be described by chemical equilibrium, it is interesting to quantify the effect of the different parameters that are likely to influence the chemical composition. \citet{Venot_2014} studied the atmospheric composition of the warm Neptune GJ 3470b. They explored the parameter space for metallicity ($\zeta$), temperature ($T$), eddy diffusion coefficient ($K_{zz}$), and stellar UV flux ($F_{\lambda}$). They varied these parameters with respect to the values of their standard model\footnote{Nominal $\zeta$ is 10 $\times$ solar metallicity and nominal $F_{\lambda}$, $T$, and $K_{zz}$ are represented on their Figs. 2 and 3 respectively.}.

\citet{Venot_2014} found that the value of the eddy diffusion coefficient and the intensity of stellar UV irradiation have a lower impact on the chemical composition, compared to the much larger effect of metallicity and temperature, which lead to changes of several orders of magnitude in the abundance of some species.

\begin{figure}[h!]
\begin{center}
\includegraphics[width=1\columnwidth]{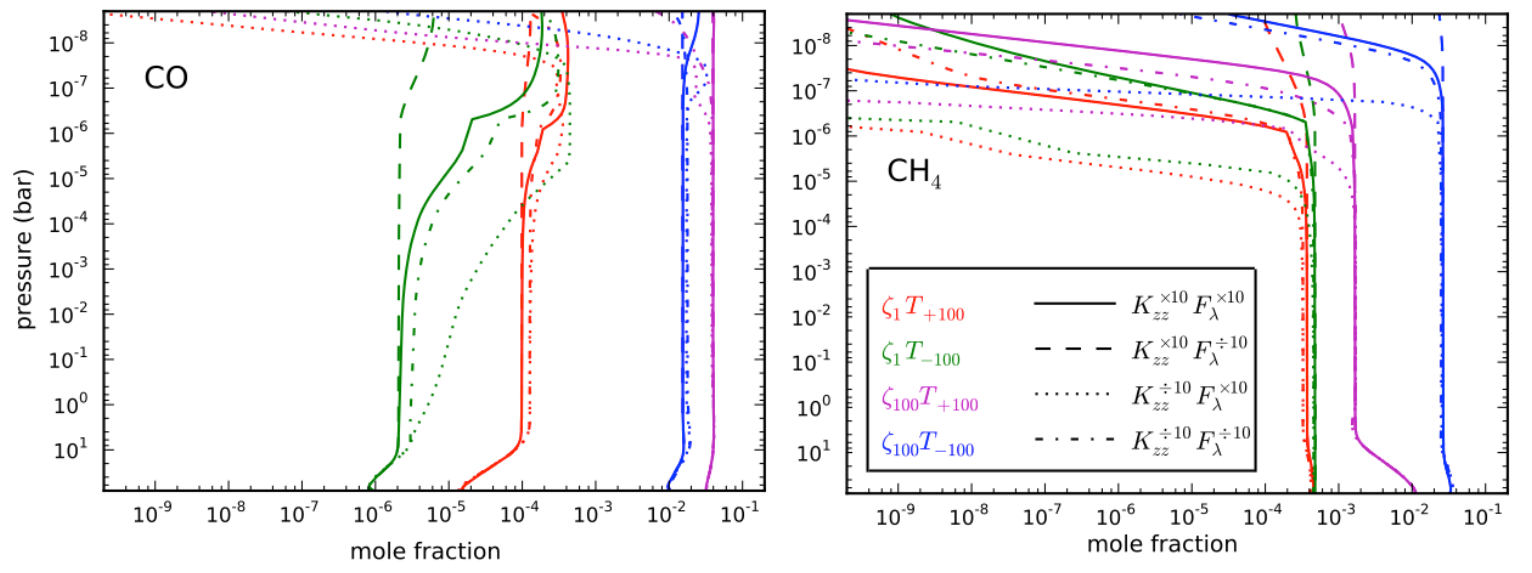}
\caption{{\label{CO_CH4_GJ3470b}
Vertical abundances profiles of CO (left) and CH$_4$ (right) from 16 models of GJ 3470b with various values of the metallicity ($\zeta$), temperature ($T$), eddy diffusion coefficient ($K_{zz}$), and stellar UV flux ($F_{\lambda}$). Each color corresponds to a set of metallicity and temperature, and each line style to a set of eddy diffusion coefficient and stellar irradiation. Adapted from \citet{Venot_2014}, reproduced with permission \copyright ESO.}}
\end{center}
\end{figure}

Fig. \ref{CO_CH4_GJ3470b} shows the vertical abundance profiles of CO and CH$_4$, the main reservoirs of carbon, for a range of values of the four parameters described above. In this case, it is apparent that metallicity and temperature are much more important in determining the mole fractions of these two species, compared with the eddy diffusion coefficient and the UV flux.

\begin{figure}[h!]
\begin{center}
\includegraphics[width=0.9\columnwidth]{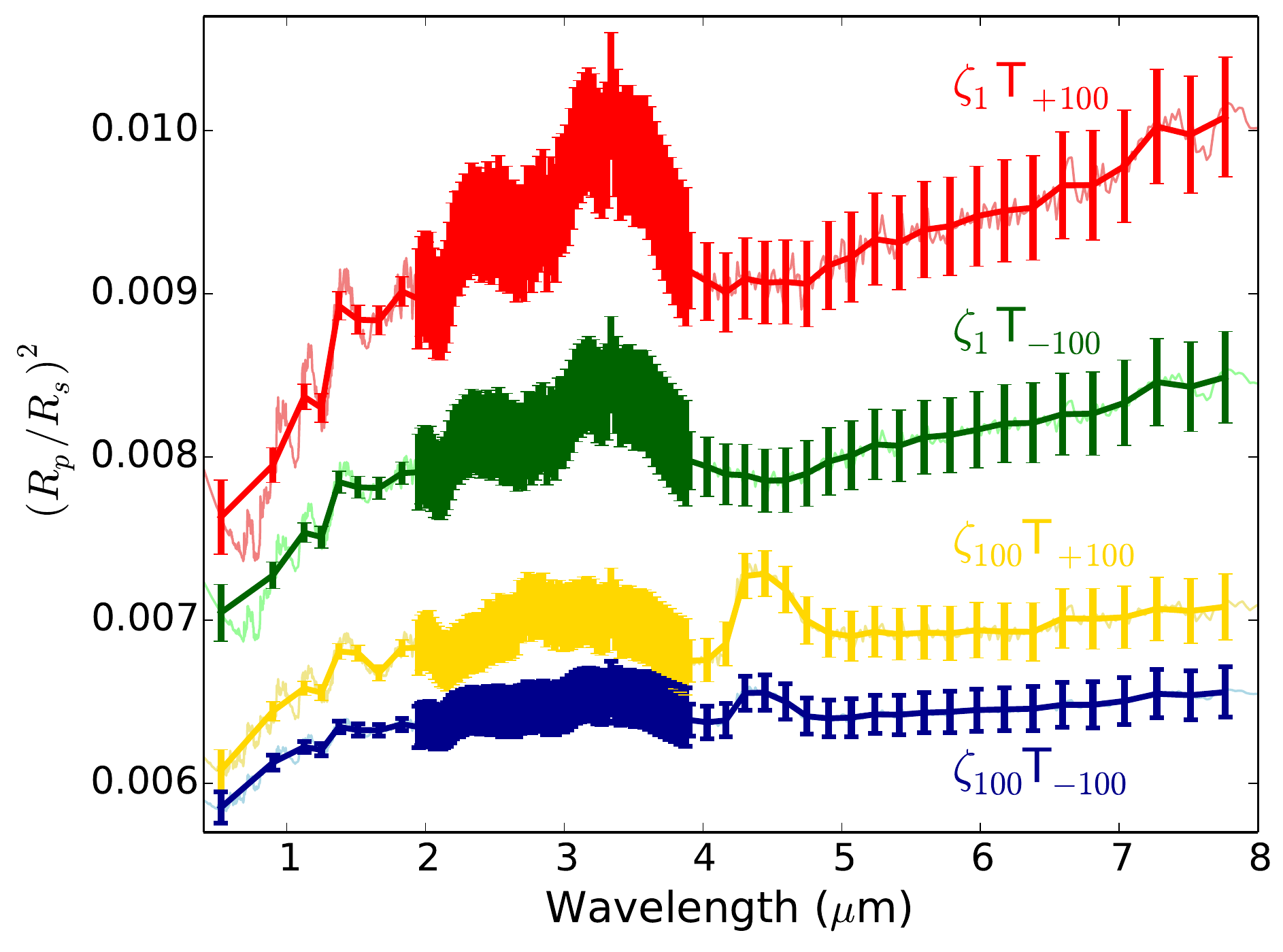}
\caption{{\label{spectra_GJ3470b_all}
Synthetic transmission spectra from 4 selected models of GJ 3470b with a high vertical mixing ($K_{zz}^{\times 10}$) and a high irradiation ($F_{\lambda}^{\times 10}$) but with different metallicities ($\zeta$) and temperatures ($T$). The meaning of the different colors is explained on each plot. Bold curves are spectra binned to ARIEL resolution, as explained in Table \ref{tab:char_ARIEL}. The error bars correspond to a SNR of 2. Fainter curves are higher-resolution spectra (R=300, constant in wavelength).}}
\end{center}
\end{figure}

Fig. \ref{spectra_GJ3470b_all} shows the transmission spectra for four parameter combinations of metallicity and temperature, each with $K_{zz}^{\times 10}$$F_{\lambda}^{\times 10}$, including simulated ARIEL observations. All of the spectra have roughly the shame shape, which is dominated by absorption features of H$_2$O and CH$_4$. The effect of changing the metallicity and temperature is to vertically shift spectra, since both parameters lead to changes in the atmospheric scale height \citep{Venot_2014}. The spectra corresponding to the different $\zeta_xT_y$ cases can be easily differentiated with a signal-to-noise ratio (SNR) of 2, corresponding to one single transit.

\begin{figure}[h!]
\begin{center}
\includegraphics[width=0.9\columnwidth]{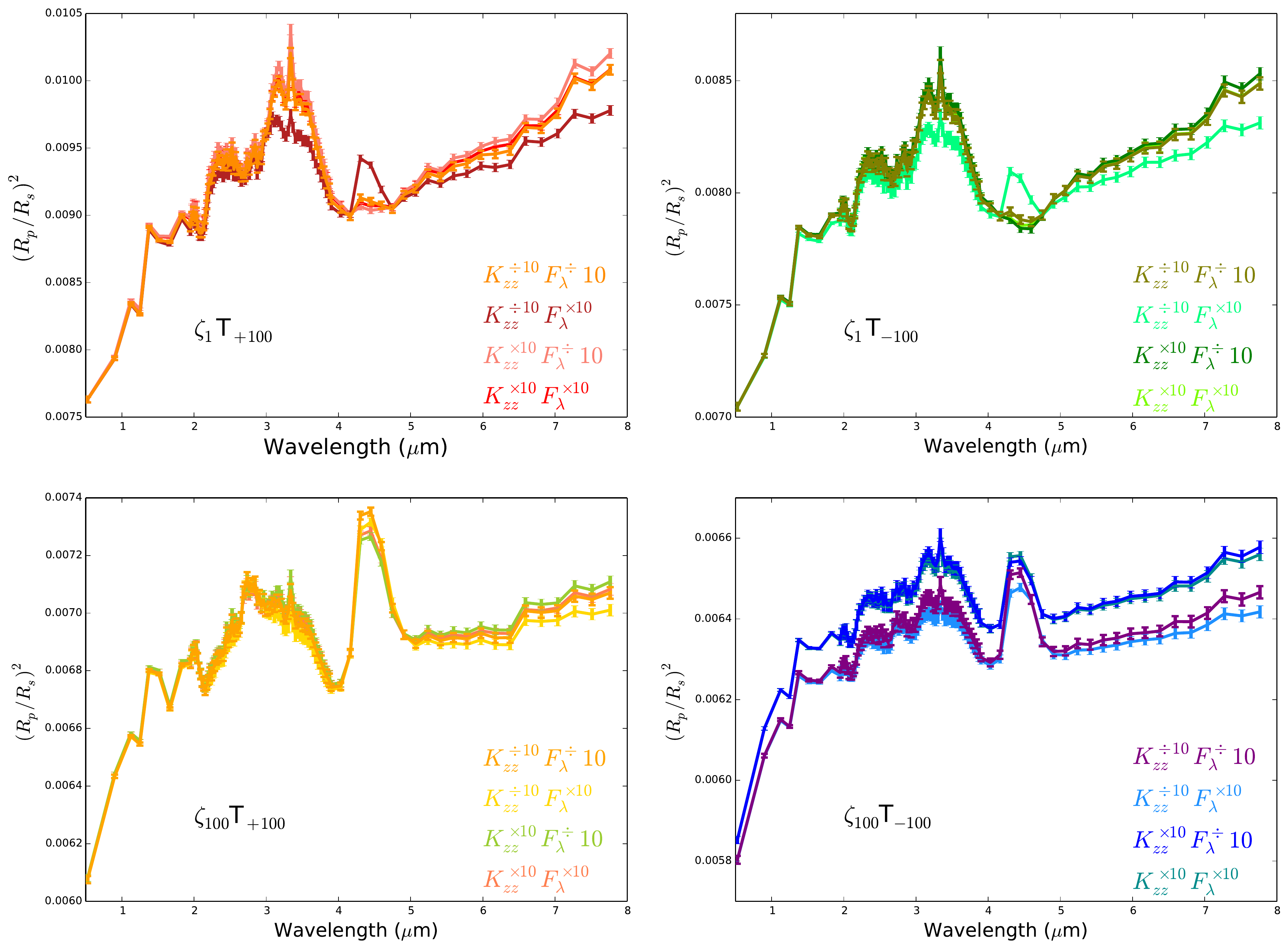}
\caption{{\label{spectra_GJ3470b}
Synthetic transmission spectra from 16 models of GJ 3470b as they will be observed by ARIEL. The resolutions are 10, 100, and 30, according to the spectral range, as explained in Table \ref{tab:char_ARIEL}. The error bars correspond to a SNR of 20. The meaning of the different colors is explained on each plot.}}
\end{center}
\end{figure}

Fig. \ref{spectra_GJ3470b} shows the effect of different eddy diffusion coefficients and UV fluxes on the transmission spectra. For a given $\zeta_xT_y$, the effect of different vertical mixing and irradiation intensities is quite moderate. For all $\zeta_xT_y$ cases, we observe variations between 5 and 8 $\mu$m, which are due to changes in H$_2$O and CH$_4$ abundances. We notice variations in the CO$_2$ signature around 4.3 $\mu$m for all the spectra of the high metallicity ($\zeta_{100}$) models. For the models with a lower metallicity (i.e. solar, $\zeta_1$), this feature appears clearly for the models with a high irradiation and a low vertical mixing ($K_{zz}^{\div10}F_{\lambda}^{\times 10}$) but is absent (or very small) for the three other ones. For these former models ($\zeta_1$ and $K_{zz}^{\div10}F_{\lambda}^{\times 10}$), we also observe a decrease of the transit depth in the CH$_4$ bands between 3 and 4 $\mu$m compared to the three other models with the same metallicity and temperature. Finally, an important variation that can be observed concerns the spectra of the $\zeta_{100}T_{-100}$ models which are shifted vertically depending on the vertical mixing. The $K_{zz}^{\times 10}$ spectra present, on the entire wavelength range, a larger transit depth than the $K_{zz}^{\div 10}$ ones. This is due to the higher atmospheric scale height of these models.
For each $\zeta_xT_y$ cases, a high SNR ($\geq$20) is necessary to differentiate the four spectra corresponding to the different $K_{zz}F_{\lambda}$ cases. Given the SNR per transit for GJ3470b (see Table \ref{tab:SNR}), a SNR of 20 is achievable by combining multiple transits.

\subsection{Eccentricity}
The effect of orbital eccentricity on the thermal profile and composition was investigated by \citet{Agundez_2014} for the eccentric warm Neptune GJ 436b. 

\begin{figure}[!hb]
\begin{center}
\includegraphics[width=0.9\columnwidth]{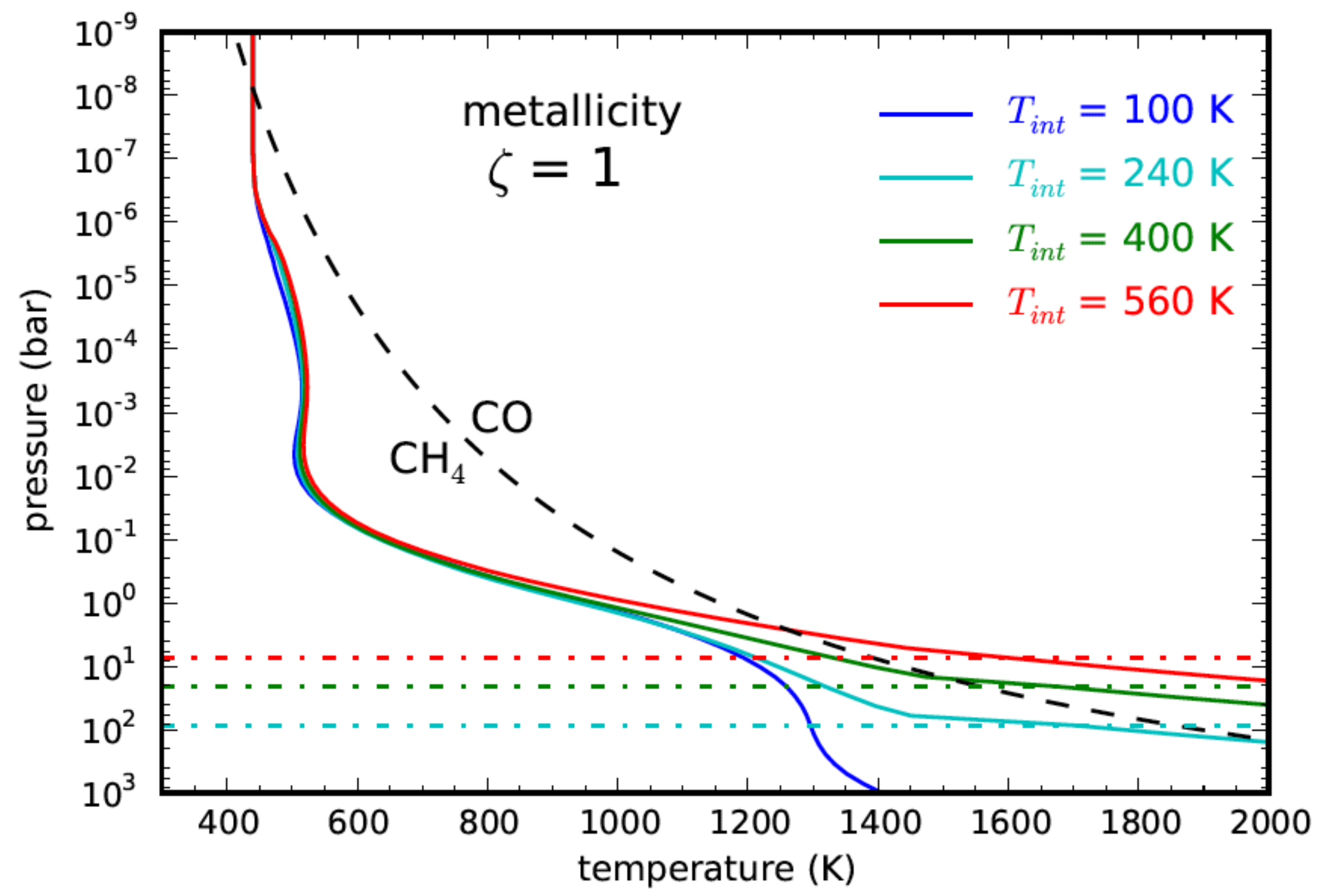}
\caption{{\label{PT_GJ436b}
Pressure-temperature profiles (full lines) of GJ~436b assuming different internal temperatures: 100 K (dark blue), 240 K (cyan), 400K (green), and 560 K (red). The dashed-dotted lines represent the transition between the radiative and the convective zone. The black dashed line represent the equilibrium line CO/CH$_4$. From \citet{Agundez_2014}.%
}}
\end{center}
\end{figure}

The dissipation of tidal forces in the planet releases an internal heat flow $\phi_{int}$. To quantify the effect of this heat flow on the thermal structure of the atmosphere, it can be incorporated into a radiative-convective model as an internal temperature $T_{int} = (\phi_{int}/\sigma)^{1/4}$, where $\sigma$ is the Stefan-Boltzmann constant. However, it is very challenging to evaluate precisely the dissipation rate of tidal forces in a planet, as it depends on the internal composition and structure, unknown in the case of GJ~436b. To study the tidal forces in GJ~436b, \cite{agu2014} used the Constant Time Lag (CTL) model with the factor $Q'$ controlling the dissipation of tidal forces. $Q'$ verifies the relation $k_2 \Delta t = 3/(2Q'n)$, where $k_2$ is the Love number of degree 2 \citep{goldreich1966}, $\Delta t$ is the time lag \citep{hut1981}, and $n$ is the orbital mean motion \citep{leconte2010}. To estimate $Q'$, they compared the age of GJ 436 system with the circularisation timescale $\tau_e = e/\dot e$ and found that $Q'$ was greater than $10^5$, given the observed eccentricity. This limit value of $Q'$ leads to an internal temperature $T_{int} < $ 560~K (see Fig. 1 of \citealt{agu2014}). For a detailed discussion on the tidal heating in exoplanet atmospheres, we refer the reader to \cite{leconte2010}.

\citet{agu2014} considered four internal temperatures that satisfy the above criterion: 100, 240, 400, and 560 K. In \cite{agu2014}, different metallicities were investigated, but here we will present only one case (i.e. 10$\times$ solar metallicity) and focus on the effect of different internal heating. The eddy diffusion coefficient is the same for each model and has been estimated from the GCM model of \cite{lewis2010} (i.e. K$_{zz}$ varying between 10$^8$ and 10$^{11}$cm$^2$s$^{-1}$ depending on the pressure level.)

\begin{figure}[!b]
\begin{center}
\includegraphics[width=0.90\columnwidth]{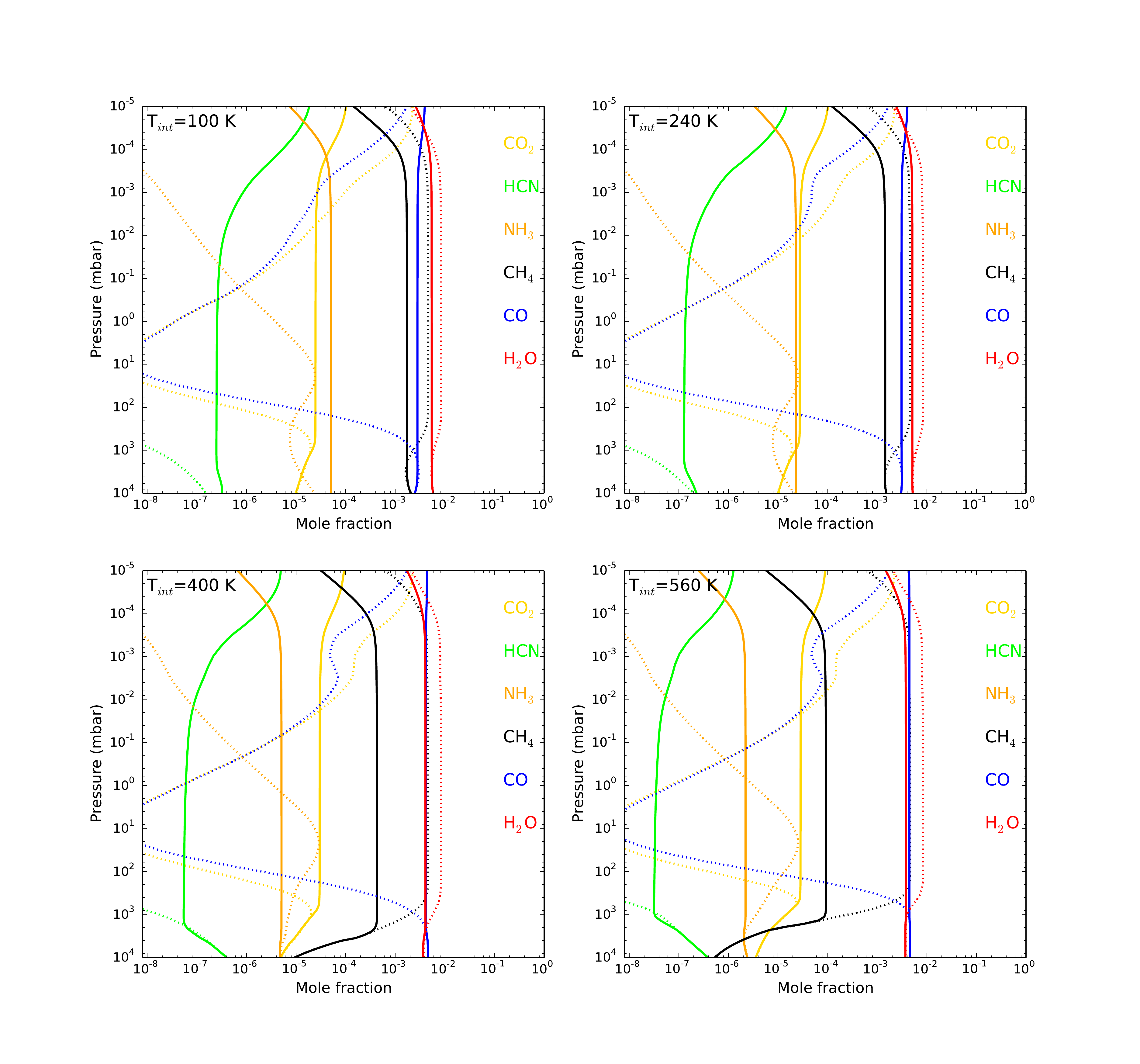}
\caption{{\label{compo_GJ436b}
Chemical composition of GJ 436b with a metallicity of 10$\times$ solar metallicity and different internal temperature (100 K, 240 K, 400 K, and 560 K) as labeled on each panel. Steady-state composition (full lines) is compared to thermochemical equilibrium (dashed lines). Adapted from \citet{Agundez_2014}.%
}}
\end{center}
\end{figure}

As can be seen in Fig. \ref{PT_GJ436b}, the various internal temperatures lead to different temperature structures, particularly in the deep atmosphere, with subsequent consequences for the chemistry. Each thermal profile crosses the CO=CH$_4$ equilibrium line at different pressure levels, that will lead to differences in the steady-state abundances. For example, if quenching happens around 10 bar, for each profile, the quenching level is located on one side or the other of this equilibrium line. 
Thus, because of quenching, the chemical composition of the atmosphere will be different. As one can see in Fig. \ref{compo_GJ436b}, the higher the internal temperature, the higher (lower) the abundance of carbon monoxide (methane) in the deep atmosphere. These variations of abundances lead to important changes regarding what is expected from thermochemical equilibrium: for all the thermal profiles with a 10$\times$ solar metallicity, CO is the major carbon-bearing species, instead of CH$_4$. For the hottest profile, CO becomes in addition the major oxygen bearing-species instead of water thanks to vertical mixing.

\begin{figure}[!b]
\begin{center}
\includegraphics[width=0.90\columnwidth]{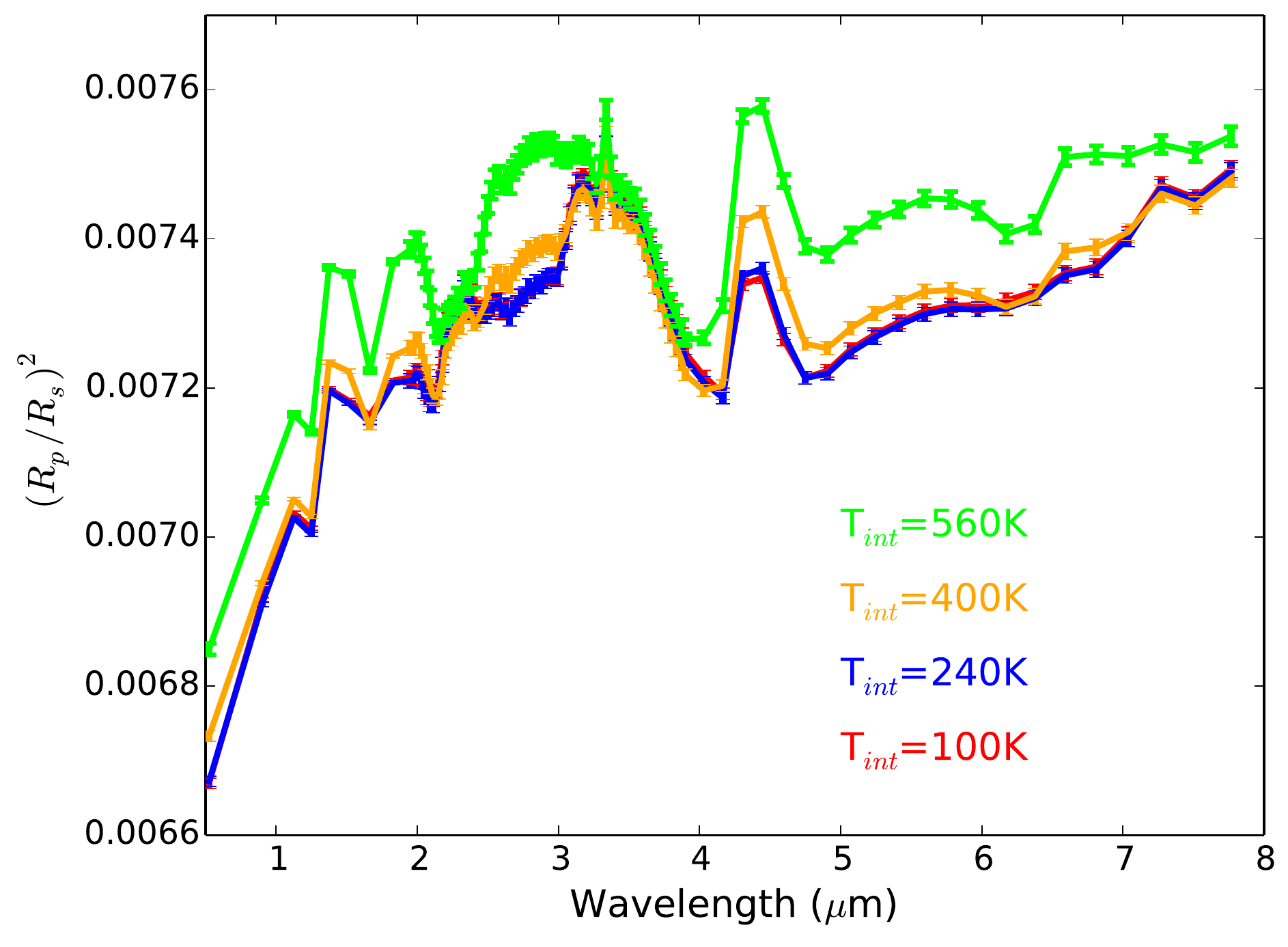}
\caption{{\label{spectra_GJ436b}
Synthetic transmission spectra of GJ 436b with a high metallicity (10$\times$ solar) and different internal temperature (100 K, 240 K, 400 K, and 560 K). Spectra are binned to ARIEL resolution, as explained in Table \ref{tab:char_ARIEL}. The error bars correspond to a SNR of 30.%
}}
\end{center}
\end{figure}

These various atmospheric compositions produce different transmission spectra (Fig. \ref{spectra_GJ436b}). To calculate the spectra, we fixed the 10 bar pressure level to the planetary radius, which we set to 0.36 R$_{Jup}$, taking into account the possible error on the value given by \cite{southworth2010} (i.e. 0.37$\pm$ 0.018). A higher radius would produce the same spectral features but with larger transit depths, more discrepant from current observational data. A complete description of the spectra and the different molecular features can be found in \cite{Agundez_2014}. We focus here on the SNR necessary to separate the different spectra.
One can see on Fig. \ref{spectra_GJ436b} that the atmosphere with the higher internal temperature is quite well separated from the other ones (between 0.8 and 230 ppm). Observations with a SNR of 10 are enough to differentiate it from the three other ones, which can be obtained combining four transits of GJ~436~b (see Table \ref{tab:SNR}). In the spectral ranges between [2.2 - 2.4] and [3.3 - 3.9] $\mu$m the four spectra are very close to each other and are therefore not differentiable, even with a very good SNR. The absorption at these ranges are due to CH$_4$. The spectrum corresponding to the model with an internal temperature of 400 K is closer to the two coldest ones with departures lower than 80 ppm. Thus, to differentiate the 400 K spectrum from the two other ones a SNR of 30 is required. Only the features around 1.4 and 4.5 $\mu$m can be separated with a lower SNR (10 or 20). To be able to draw high-confidence conclusions, a higher SNR is advisable. Finally, the spectra corresponding to the two coldest thermal profiles are too similar and not differentiable whatever the SNR of the acquisitions.

\subsection{Elemental carbon/oxygen ratio}

\begin{figure}[hbt]
\begin{center}
\includegraphics[width=0.49\columnwidth]{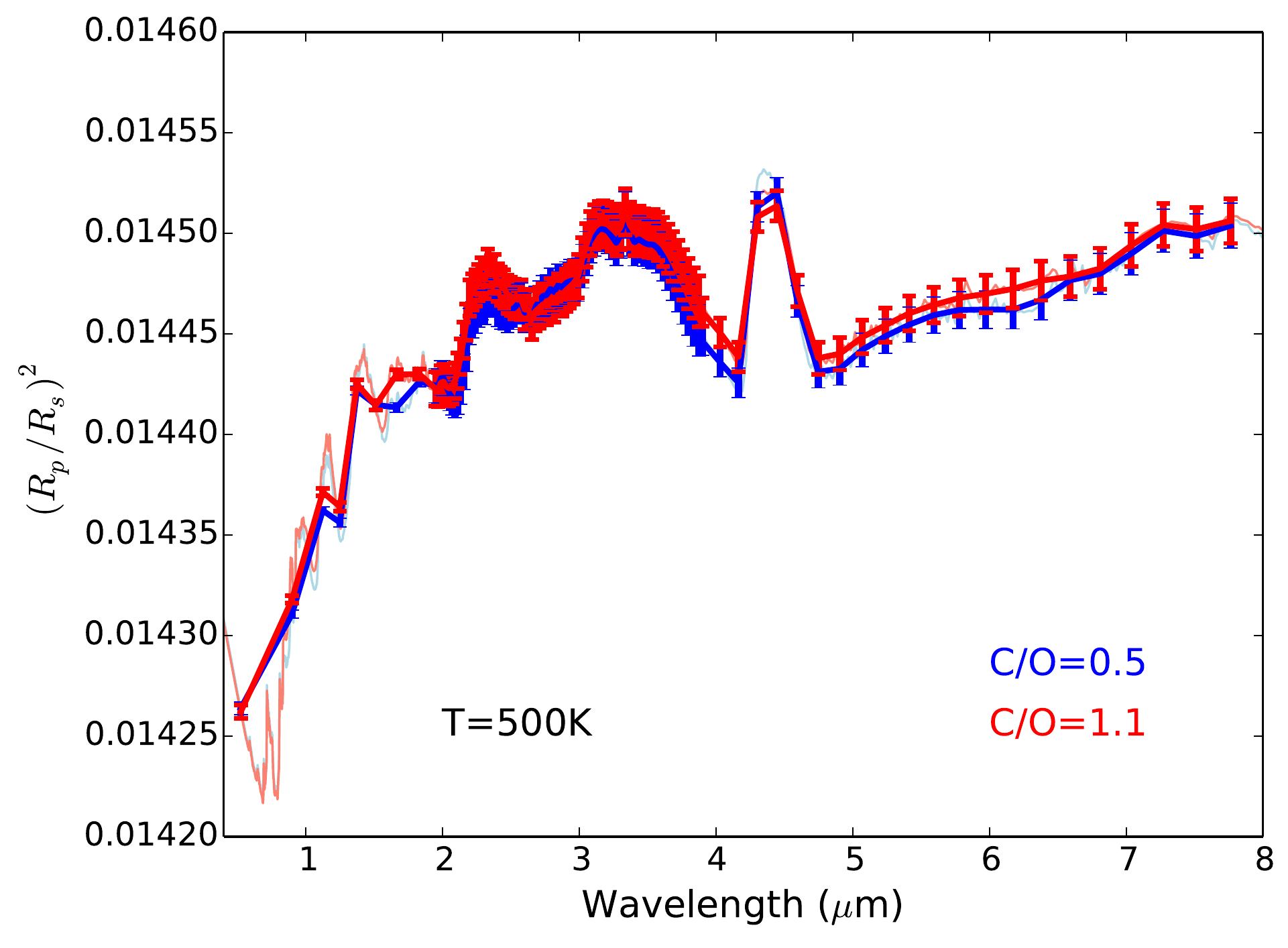}
\includegraphics[width=0.49\columnwidth]{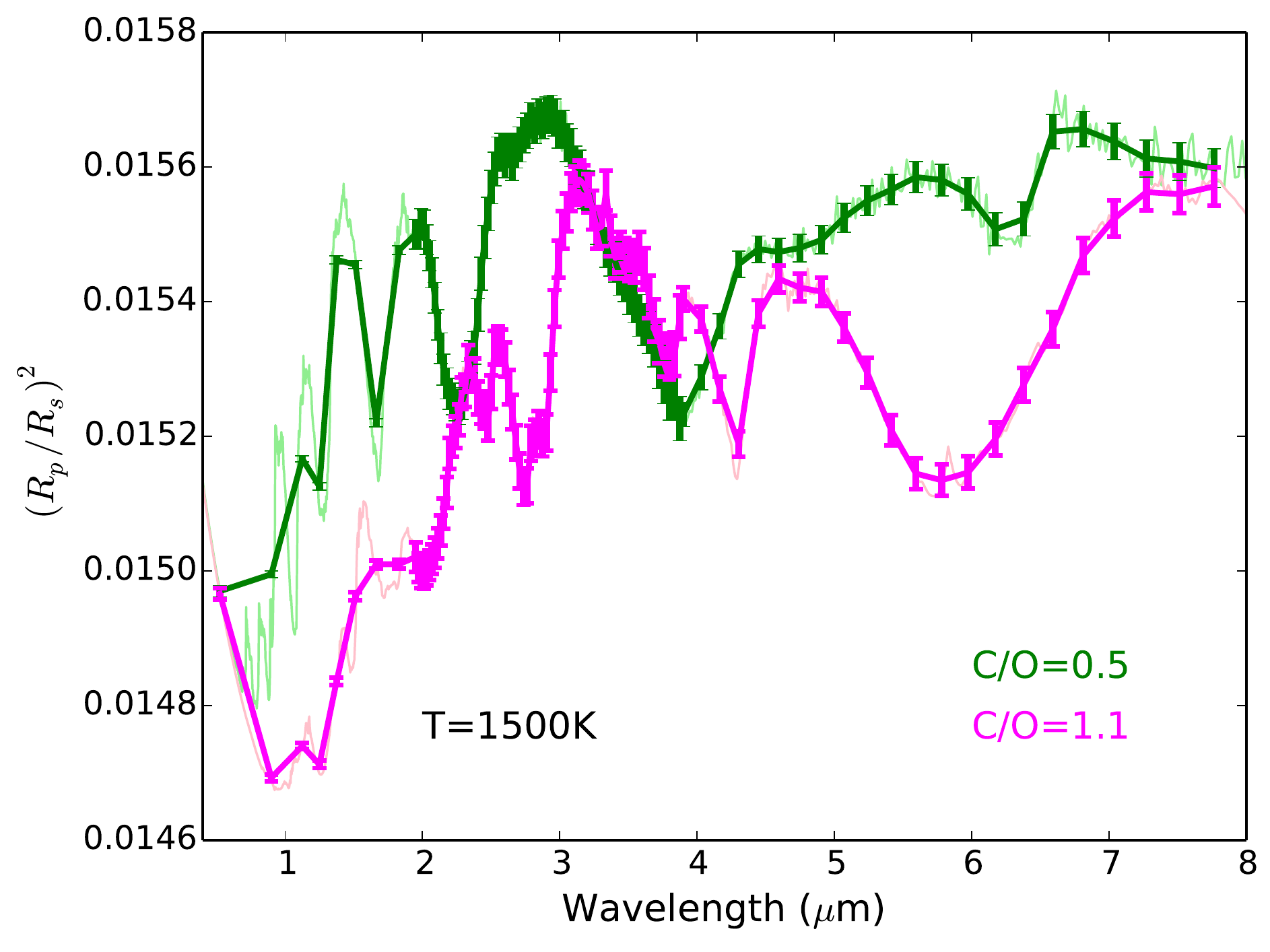}
\caption{{\label{spectra_500K_1500K}Synthetic transmission spectra for an exoplanetary atmosphere with a temperature around 500K (left) and 1500K (right). Two C/O ratios are represented: 0.5 (blue/green) and 1.1 (red/magenta). We assumed the test-planet was orbiting a Sun-like star and took the SNR of HD~189733b. The bold and low-resolution spectra are simulated ARIEL observations. The resolutions are 10, 100, and 30, according to the spectral range, as explained in Table \ref{tab:char_ARIEL}. The error bars correspond to a SNR of 20. The fainter curves are the corresponding higher-resolution spectra (R=300).}}
\end{center}
\end{figure}

Elemental abundances can have a crucial effect on the atmospheric composition of exoplanets. Several studies have been undertaken on this subject such as \cite{line2010,mad2011,mad2012,kopparapu2012,moses2013,Venot_2015,molliere2015,heng2016,rocchetto2016} in which they also studied the consequences of different C/O ratios on the observations. Most of these studies focused on hot Jupiters and revealed that as the C/O ratio increases, abundances of hydrocarbons species increase also. \cite{Venot_2015} studied less hot exoplanets and found that for atmospheres with a temperature around 500 K, changing the C/O ratio from solar (C/O = 0.54) to twice solar (C/O = 1.1) has only a minor effect on the chemical composition and on the synthetic spectra (see Fig. \ref{spectra_500K_1500K}). These two synthetic spectra would be very difficult to differentiate, even with a very good SNR.

\begin{figure}[!ht]
\begin{center}
\includegraphics[width=0.8\columnwidth]{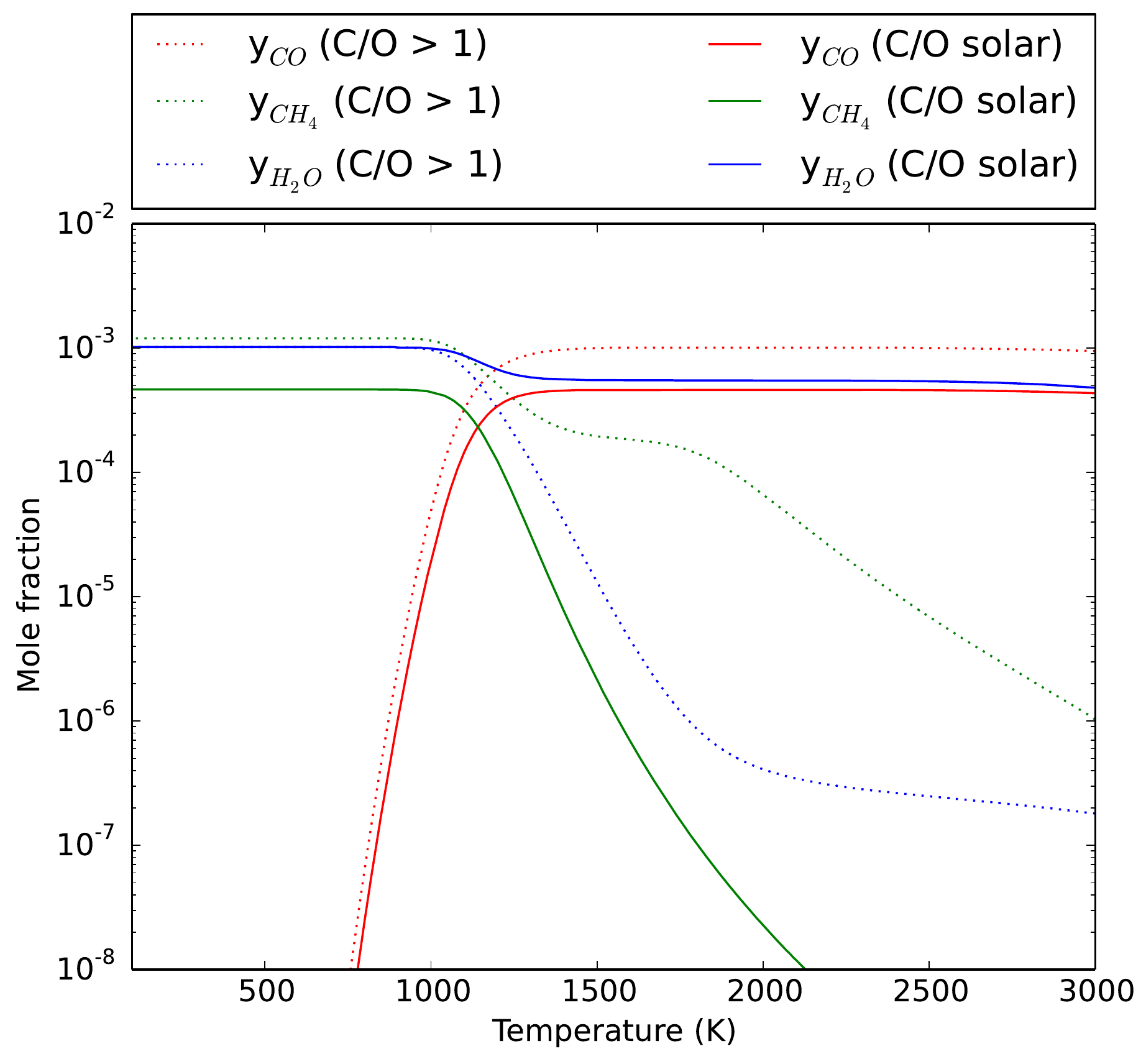}
\caption{{\label{abundances_thermo}Abundances of CO (red), CH$_4$ (green), and H$_2$O (blue) at thermochemical equilibrium as a function of temperature. Calculations have been done for P = 1 bar and for two Carbon/Oxygen ratio: solar (full lines) and $>$ 1 (dotted lines).}}
\end{center}
\end{figure}

Indeed, at this temperature, thermochemical equilibrium predicts that the most abundant heavy species remain the same at C/O solar or above 1, i.e. H$_2$O and CH$_4$ (Fig. \ref{abundances_thermo}). Whereas the abundance of H$_2$O does not change, CH$_4$ sees its abundance doubling and becoming slightly above that of water. For both C/O ratios, at 500 K, CO has a very low abundance ($\sim$10$^{-17}$ at 1 bar). Consequently, as the main species (after H$_2$ and He) are H$_2$O and CH$_4$ whatever the C/O ratio at this temperature, chemical kinetics is governed by the same reactions, leading to similar atmospheric composition.

\begin{figure}[h!]
\begin{center}
\includegraphics[width=0.9\columnwidth]{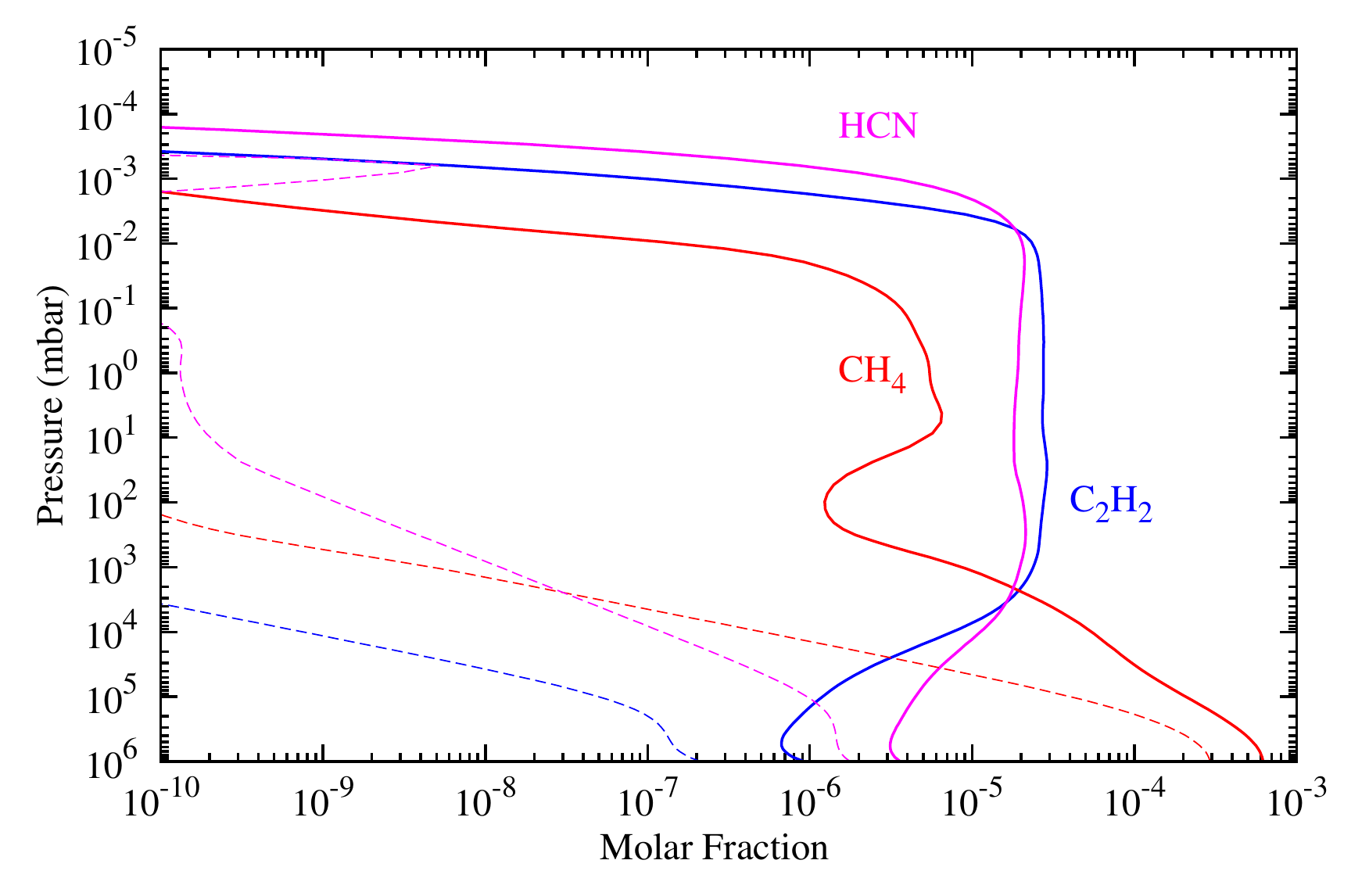}
\caption{{\label{abundances_1500K}
Vertical abundance profiles of CH$_4$ (red), C$_2$H$_2$ (blue), and HCN (pink) for an atmosphere with a temperature around 1500 K assuming two different C/O ratios: solar (dashed lines) and twice solar (full lines). Adapted from \citet{Venot_2015}, reproduced with permission \copyright ESO.%
}}
\end{center}
\end{figure}

In agreement with previous studies aforementioned, \cite{Venot_2015} found that when dealing with hotter planets, the effect of the elemental abundances is much more important. The increase of the C/O ratio leads to an important increase (by several orders of magnitude) in the abundance of hydrocarbon species (i.e. CH$_4$, C$_2$H$_2$, etc.), accompanied with a decrease of the abundance of water (see Fig. \ref{abundances_1500K}). 
Here again, thermochemical equilibrium calculations provide an explanation for these results. Figure \ref{abundances_thermo} shows that at T = 1500 K, contrary to T $\lesssim$ 1000 K, the major species change with the C/O ratio. For a solar value, H$_2$O and CO are the main heavy species and CH$_4$ is less abundant by more than two orders of magnitude. On the contrary, for a C/O ratio above 1, the abundance of water is much lower and the major species are CO and CH$_4$. Consequently, the main reactions and destruction/formation pathways in a hot atmosphere will be different depending on the C/O ratio. See also \cite{moses2013} for a detailed study on the effect of the C/O ratio on hot exoplanetary atmospheres.

This difference of chemical composition is highly visible on the synthetic spectra. As one can see in Fig. \ref{spectra_500K_1500K}, the two transmission spectra corresponding to the C-rich and the C/O solar cases are very separate and can be differentiated easily with a low SNR (5). The shape of the spectrum corresponding to the C/O solar case is mainly due to water absorption, with some features of CO and CO$_2$ around 4.8 $\mu$m. In contrast, the C-rich spectrum owes its form to CH$_4$, CO, HCN, and C$_2$H$_2$. Because of their strong spectral features around 14 $\mu$m, these latter two absorbers can be used as tracers for the C/O ratio in warm exoplanet's atmosphere \citep[i.e.][]{kopparapu2012,Venot_2015}. This technique has been used by \citet{Tsiaras_2016} to suggest that the super-Earth 55 Cancri e possesses a C-rich atmosphere, thanks to the detection of high amount of HCN.

As it has been shown by \cite{rocchetto2016}, differentiating an atmosphere with a C/O $<$ 1 from a C/O $>$ 1 is quite straightforward with a retrieval process as the spectra are really dissimilar but it might be more difficult to retrieve a more precise value of the C/O ratio. Indeed, an atmosphere with a C/O = 0.5 has a spectrum very close to that of an atmosphere with a C/O = 0.6 for instance.

\section{Influence of stellar parameters}\label{sec:star_param}
\subsection{Spectral type of stars}
A crucial factor for understanding exoplanet atmospheres is the host star. Stellar irradiation largely determines the temperature of the atmosphere and, in addition, the stellar flux is observable in the spectra through reflected and transmitted light. Furthermore, UV flux irradiation drives photochemistry in the upper atmospheres of exoplanets.

\begin{figure}[h!]
\begin{center}
\includegraphics[width=0.9\columnwidth]{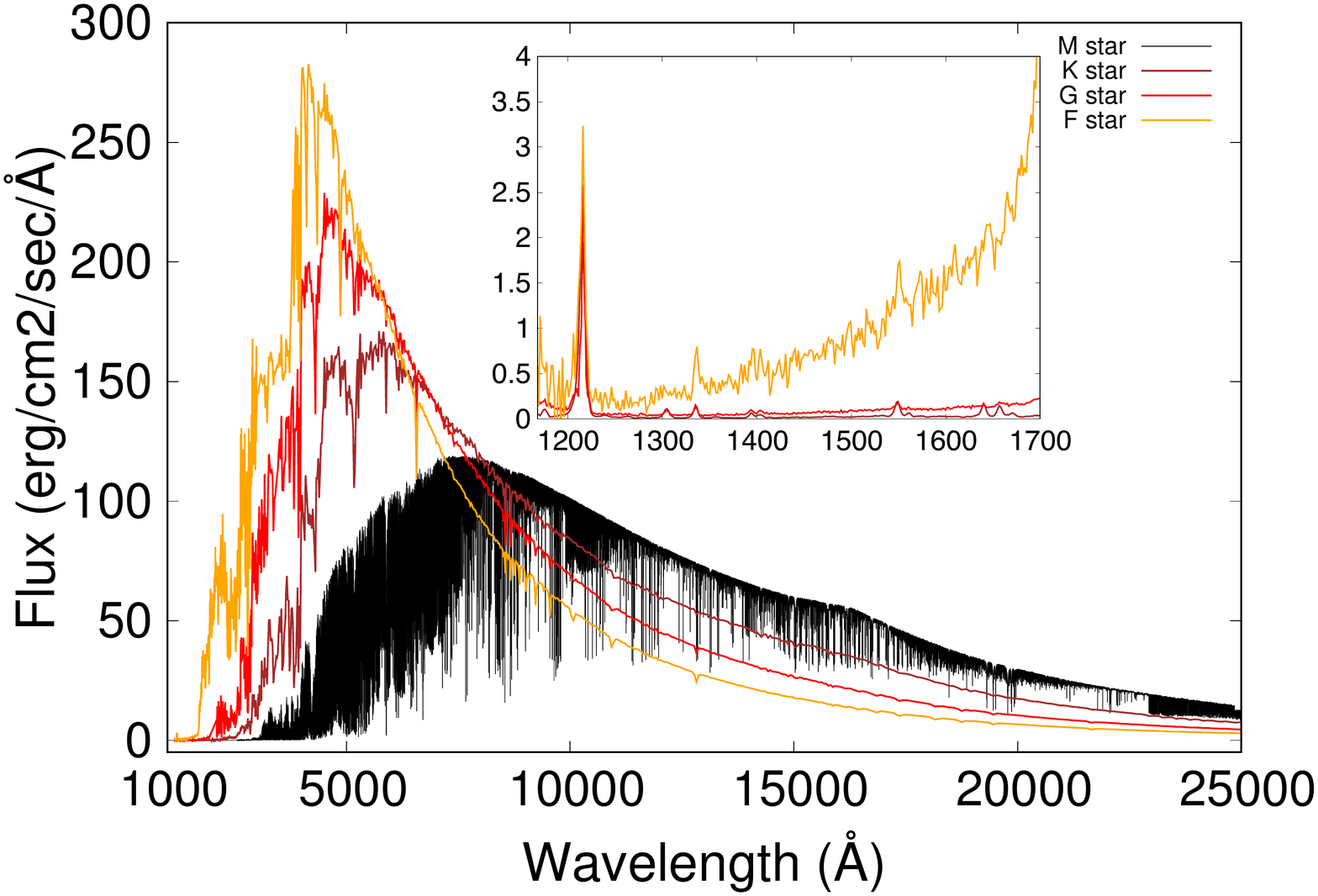}
\caption{{\label{stellar-flux}
Stellar flux at 1 AU as a function of wavelength of an M (black), K (brown), G (red) and F (orange) star. FarUV flux can be seen in the zoomed region.}}
\end{center}
\end{figure}

Fig. \ref{stellar-flux} shows the stellar spectra of an M, K, G and F star. These spectra are a combination of data from the ATLAS synthetic spectra \citep{kurucz1979} and UV observations from the International Ultraviolet Explorer (IUE) as explained in \citet{ru13}. For the M star, we are showing the emission flux of an inactive M star from a model by \citet{Allard2001}. As seen in the figure, stars of different stellar types emit very different fluxes, creating a different response in the atmospheres of the planets around them.

\begin{figure}[h!]
\begin{center}
\includegraphics[width=0.9\columnwidth]{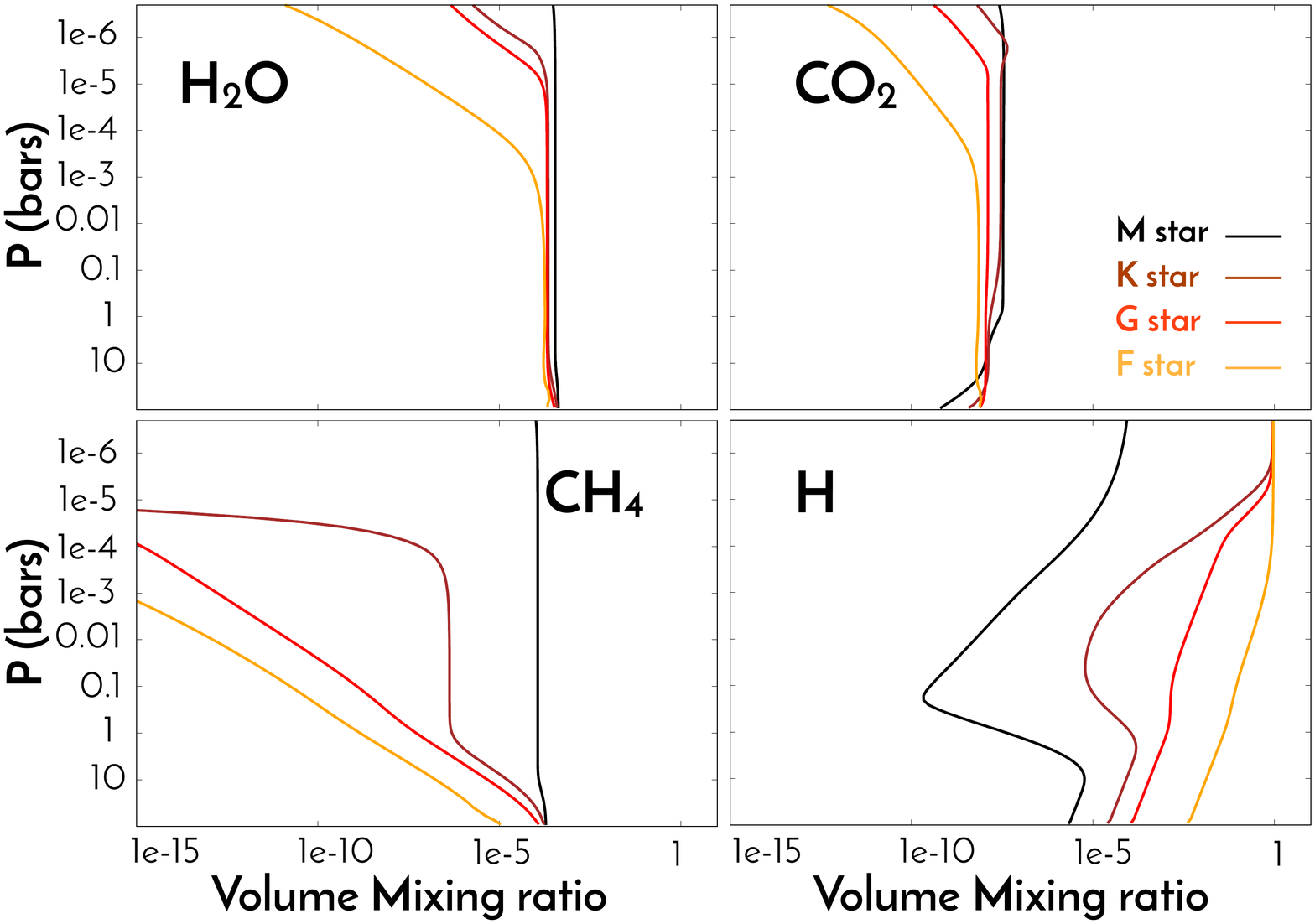}
\caption{{\label{stellartypes}
Vertical chemical abundances profiles of H$_2$O, CO$_2$, CH$_4$, and H in an exoplanet atmosphere when the planet orbits different stellar type stars. The different stellar types (F, G, K, M) are represented by different colors.}}
\end{center}
\end{figure}

\citet{miguel2014} studied the effect of stellar flux on the chemistry of extrasolar giant planets and mini-Neptunes. They studied the change in atmospheric chemical composition for planets around different stars and semi-major axis ranging from 0.01 to 0.1 AU. In Fig. \ref{stellartypes} we show how the atmospheric composition of an exoplanet is affected when the planet is exposed to irradiation from different stellar types stars. The simulations were performed for a planet located a 0.025 AU and exposed to the irradiation of the stars shown in Fig. \ref{stellar-flux}. The TP profiles for each planet were calculated using a gray atmosphere \citep{guillot2010}, as explained in \citet{miguel2014}. The results show that the stellar flux (especially in the UV domain) strongly influences the photochemistry, especially the photolysis of H$_2$O. Since water drives the chemistry in the region between 10$^{-4}$ and 10$^{-6}$ bars of these planet's atmospheres, the change in water affects the chemistry of the other major species. A planet around an F star receives a much higher amount of UV flux than a planet around an M star, therefore photolysis of water is much more efficient for the planets around F and G stars than it is on planets around cool K and M stars. When looking at pressures around 10$^{-4}$ bars, we see that the change in H$_2$O and CO$_2$ between the two extreme cases (planet around M and F stars) is of approximately two orders of magnitude, and the change in composition of CH$_4$ and H is substantial. The variation of CH$_4$ abundance between the two extreme values at that pressure is of approximately 10 orders of magnitude and the variation in H mixing ratio is of 5 orders of magnitude. 

\begin{figure}[h!]
\begin{center}
\includegraphics[width=0.9\columnwidth]{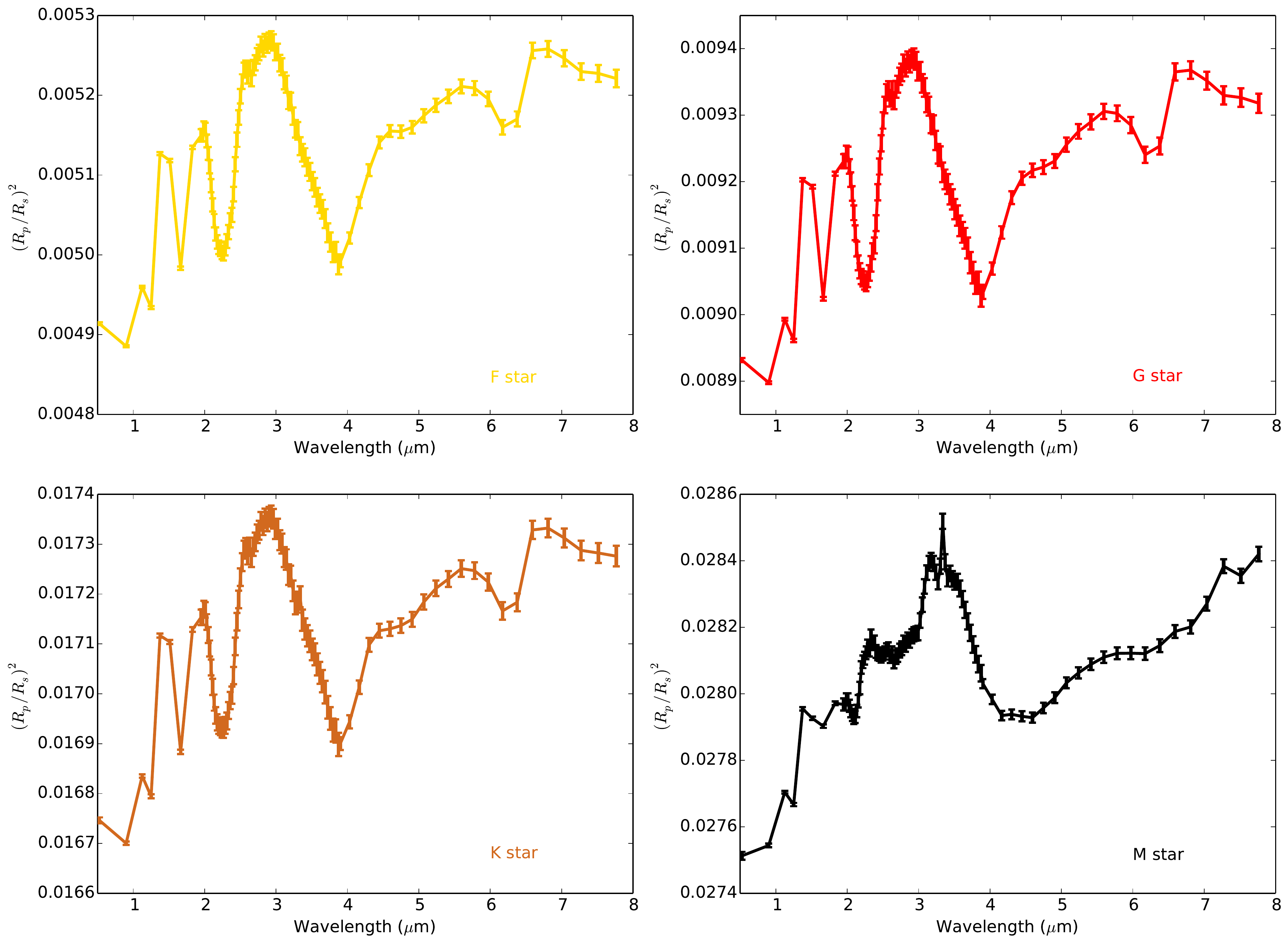}
\caption{{\label{stellartypes-spec}
Synthetic transmission spectra of a Jupiter-like planet orbiting around an F, G, K, and M star as it will be observed by ARIEL. The resolutions are 10, 100, and 30, according to the spectral range, as explained in Table \ref{tab:char_ARIEL}. The error bars correspond to a SNR of 30.}}
\end{center}
\end{figure}

We calculated the transmission spectra for the four models presented in Fig. \ref{stellartypes}. The four spectra present very different transit depth because the F, G, K, and M stars have different radii (1.5 $R_\odot$, 1.1 $R_\odot$, 0.8 $R_\odot$, and 0.62 $R_\odot$ respectively) while the planet keeps the same properties (i.e. $R=R_{Jup}$ and $M=0.969~M_{Jup}$). Apart from this obvious difference, we notice that the spectra for the F, G, and K host stars present a similar shape. All the absorption is due to water, except the absorption feature around 4.8 $\mu$m, which is the contribution of CO$_2$ and CO. The spectrum of the planet orbiting a M star presents all the absorption features of H$_2$O (i.e. around 0.95, 1.15, 1.4, 1.9, and 6.7 $\mu$m and on the ranges [2.5--3] and [4.2--5.8] $\mu$m) to which are superimposed those of CH$_4$ around 1.6 and 2.3 $\mu$m and on the ranges [3--4] and [7--8] $\mu$m. The absorption features of these two molecules will be easily characterized by ARIEL with observations with a SNR $\geq$15.

The effect of stellar flux is extremely important for hot and warm exoplanet's atmospheres, but it can also affect smaller planets at wider orbits. The main effect of stellar irradiation in rocky planet's atmospheres located in the habitable zone \citep{ko13} is that as temperature increases (going from cold K stars to hot F stars) the atmospheres show more O$_3$, more OH, less tropospheric H$_2$O (but more stratospheric H$_2$O), and less stratospheric CH$_4$, N$_2$O, and CH$_3$Cl, showing that the effect of the star is extremely important for the modelling and thus the understanding of their atmospheres \citep{ru13}. 

\subsection{Stellar activity}
\begin{figure}[!hbt]
\begin{center}
\includegraphics[width=0.8\columnwidth]{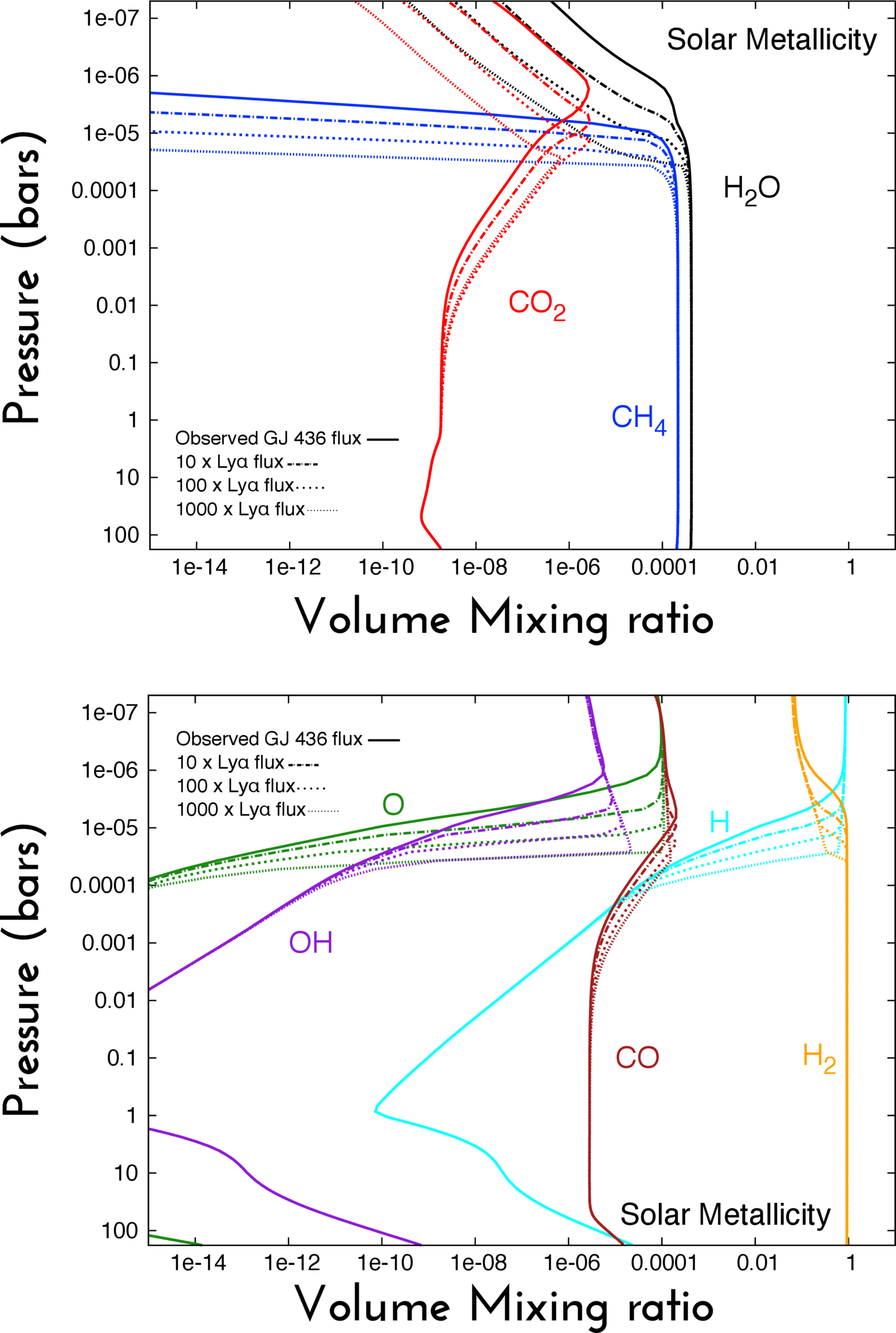}
\caption{{\label{fig:effectUVflux}
Effect of UV flux emitted by M stars (specifically Lyman $\alpha$ radiation) on an exoplanet atmosphere. In this case the planet used as an example is the mini-Neptune GJ 436b. The figure shows the results of adopting different levels of Lyman $\alpha$ flux on the abundance of key species in its atmosphere when using a solar composition. Figure adapted from \citet{miguel2015}.%
}}
\end{center}
\end{figure}

M stars are a special case. These stars make up the vast majority of stars in the Galaxy and we expect to find many more planets around M stars in the near future. The future telescope ARIEL will allow us to characterize their atmospheres, so it is important to anticipate their possible compositions. M stars are cool stars, but not very quiet: they present a high photospheric activity that causes an excess in the UV flux produced when comparing to the black body radiation. They can also present flares that might affect the chemistry in their atmospheres, as explained in Section \ref{sec:flares}.

\subsubsection{Lyman $\alpha$}

\begin{figure}[!b]
\begin{center}
\includegraphics[width=0.9\columnwidth]{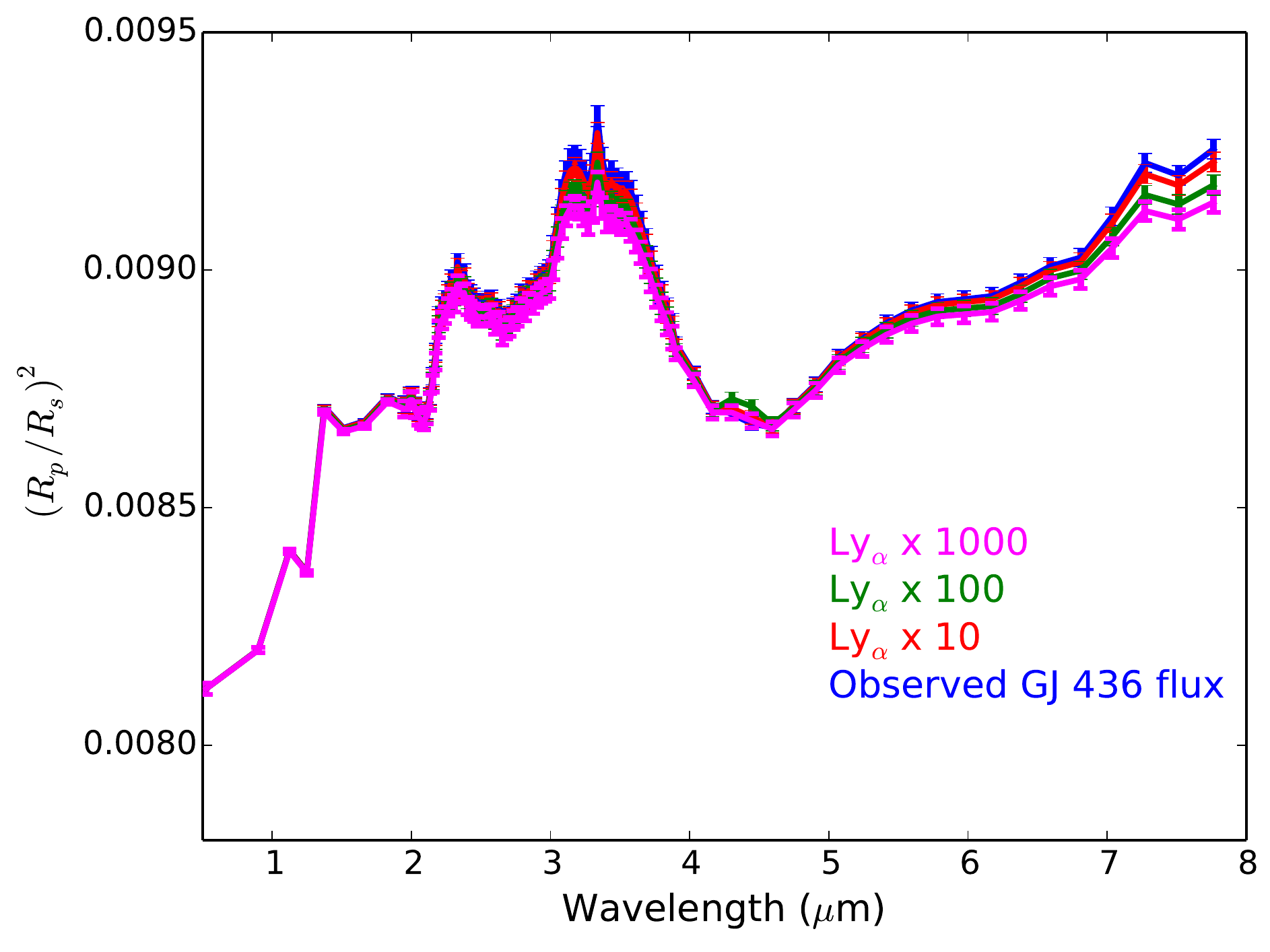}
\caption{{\label{spectra_lya}
Synthetic transmission spectra of the mini-Neptune GJ 436b when receiving different levels of Lyman $\alpha$ flux. Spectra are binned to ARIEL resolution, as explained in Table \ref{tab:char_ARIEL}. The error bars correspond to a SNR of 30.%
}}
\end{center}
\end{figure}
The brightest emission line in the UV of M stars is the Lyman $\alpha$ radiation at 1216.67 Ang, with a percentage of total UV flux from the star in the Lyman $\alpha$ line between 37 and 75 per cent compared to 0.04 per cent for the Sun \citep{France2013}. A number of M stars were observed in the UV thanks to the efforts of the MUSCLES team \citep{MUSCLESI}, and the Lyman $\alpha$ lines of those stars were reconstructed \citep{MUSCLESII}.

\citet{miguel2015} studied the effect of incoming Lyman $\alpha$ radiation on the photochemistry of mini-Neptunes' atmospheres made of solar and higher metallicities. They studied the effect of a star with different levels of Lyman $\alpha$ flux and its effect on the chemistry. In the solar metallicity atmosphere, their results show that H$_2$O presents the largest change as it absorbs most of the radiation, shielding other molecules. It can be seen on Fig. \ref{fig:effectUVflux} that H$_2$O dissociates very efficiently and the products of that dissociation (O and OH) affect the chemistry of other species \citep{moses2011}. For higher metallicities, CO$_2$ is also highly effected.

We calculated the synthetic transmission spectra corresponding to these different cases and binned them to ARIEL resolution (Fig. \ref{spectra_lya}). We observe variations around 2.5 $\mu$m and 3.5 $\mu$m and on the range [7--8] $\mu$m, which are due to changes in CH$_4$ abundance whereas the variations around 4.3 $\mu$m correspond to the increase of CO$_2$ abundance with metallicity. These  spectral variations are rather small but could be detected at some wavelength, with a very good SNR. In the case of GJ 436b, a SNR $\geq$ 30 is necessary.
These results further show that we need to obtain good observations of the stellar fluxes, especially in the UV to get a proper interpretation of the planetary spectra.

\subsubsection{Flares}\label{sec:flares}
\begin{figure}[!b]
\begin{center}
\includegraphics[width=0.9\columnwidth]{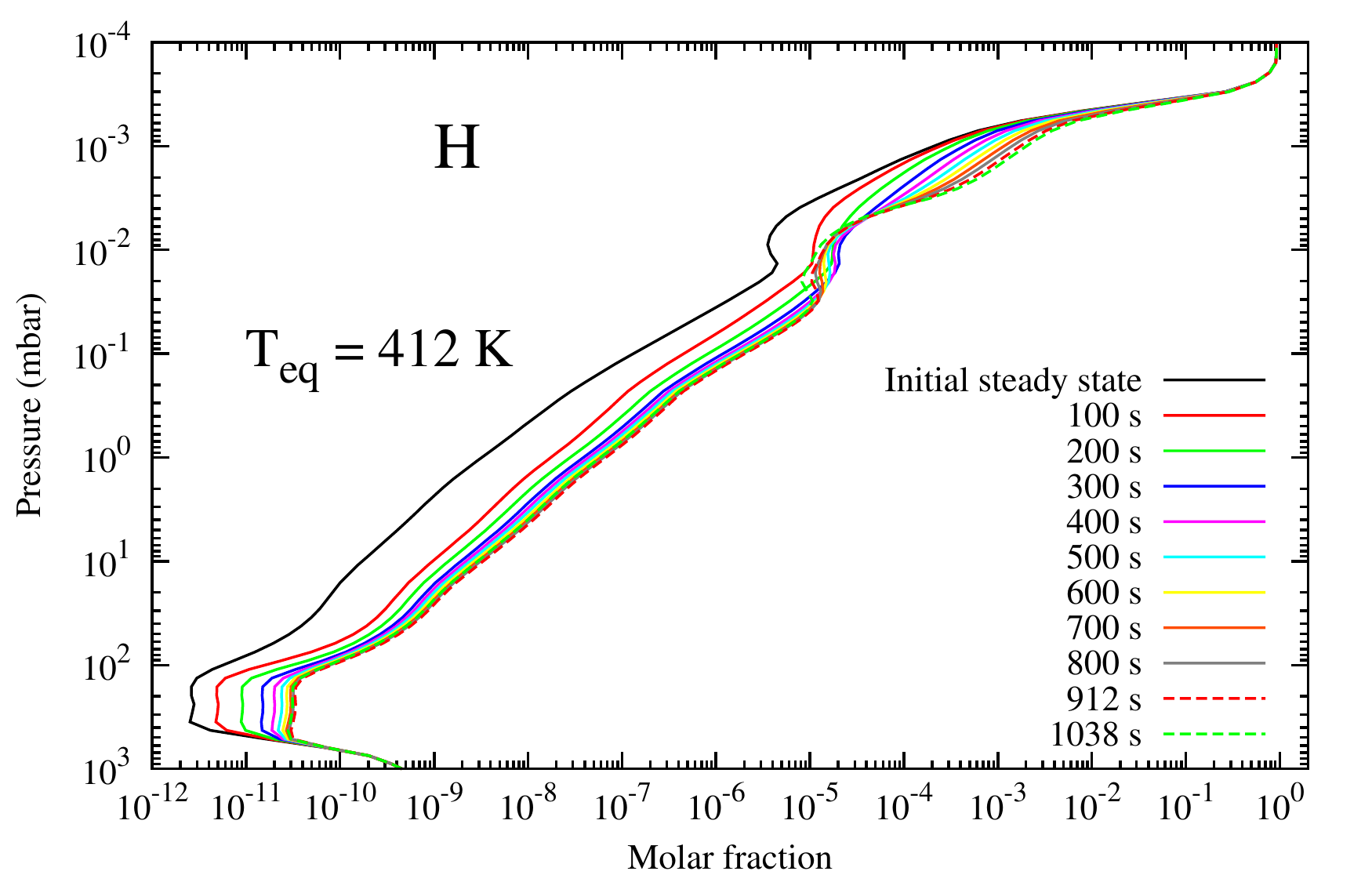}
\caption{{\label{abH_flare}
Relative abundances of hydrogen during the first 1038 seconds of a flare event. The planet is a warm Neptune with an equilibrium temperature of 412 K. From \citet{Venot_2016}.%
}}
\end{center}
\end{figure}
Active stars, in particular M stars, are subject to stellar variations, which may have an impact on the chemical composition of exoplanets. \citet{Venot_2016} studied to what extent a stellar flare can influence the atmospheric composition of hot/warm exoplanets and the resulting spectra.
They found that the increase of UV flux associated to a single flare event from an M star can modify the abundances of the main species (H, NH$_3$, CO$_2$, etc.) by several orders of magnitude and down to a pressure level of $\sim$1 bar (Fig. \ref{abH_flare}).

\begin{figure}[!tb]
\begin{center}
\includegraphics[width=0.9\columnwidth]{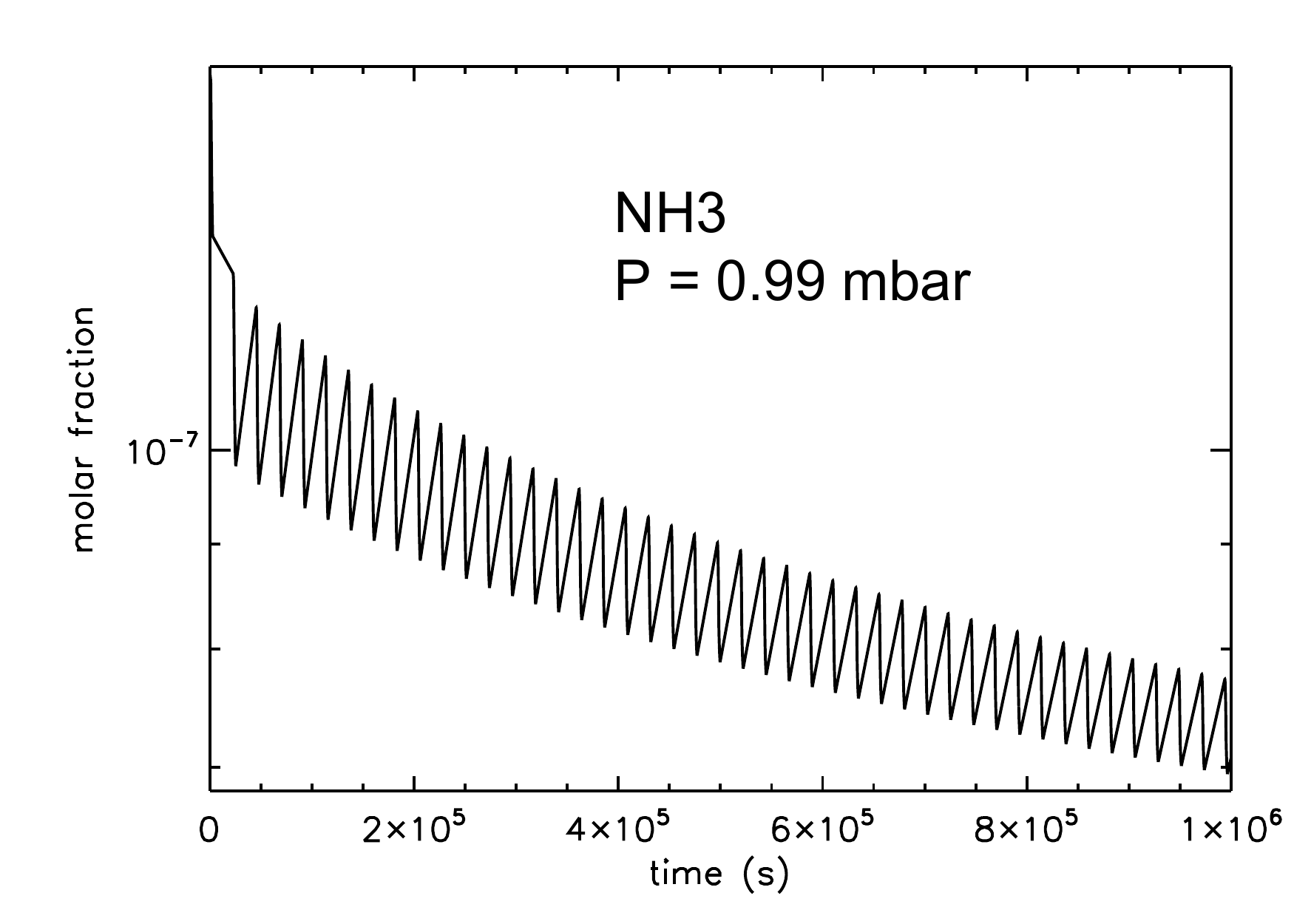}
\caption{{\label{abNH3_flare}
Temporal evolution of the atmospheric abundance of ammonia in a planet orbiting a flaring star. The host star undergoes energetic flares every five hours. The planet is a warm Neptune with an equilibrium temperature of 1303 K. From \citet{Venot_2016}.%
}}
\end{center}
\end{figure}

\begin{figure}[h!]
\begin{center}
\includegraphics[width=0.9\columnwidth]{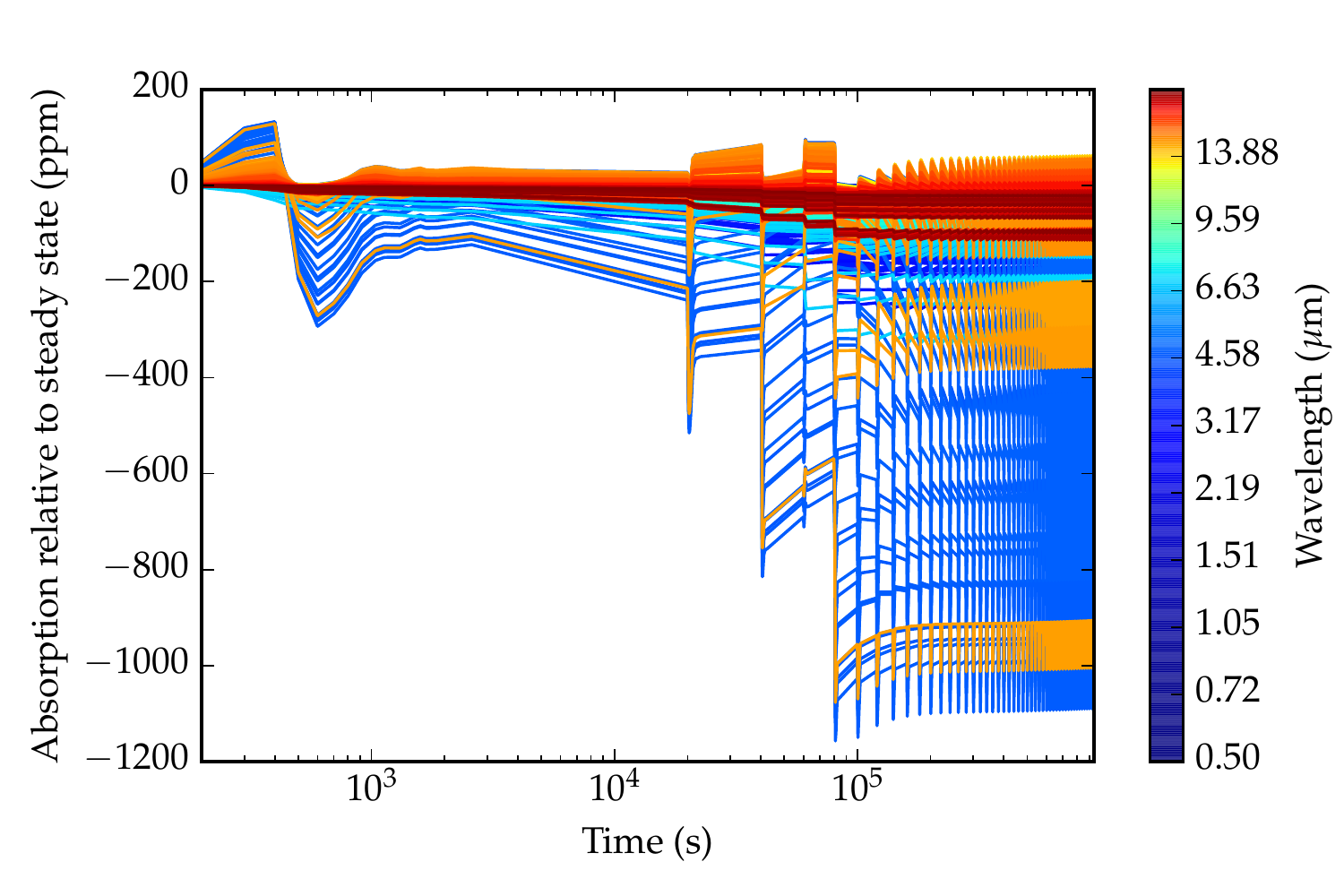}
\caption{{\label{spectrum_recu}
Variations of the spectra as a function of time and wavelength (color bar) with respect to the steady-state when the star undergoes a flare every 5 hours. From \citet{Venot_2016}.%
}}
\end{center}
\end{figure}

However, a unique flare event is not realistic, as an active star undergoes flare several time in its lifetime.
To model a more realistic case, they simulated a series of flares occurring periodically (every five hours) and found that the chemical abundances of species oscillate around a mean value that evolves with time towards a limiting value (Fig. \ref{abNH3_flare}). The number of flare events required to reach a limiting value depends on the species and on pressure. Here, the flare is happening with the same intensity and comes back with the same period. The mean and limiting values might be more difficult to reach in reality, with flares of different intensity happening with inconstant period.
The conclusion that can be drawn from this study is that planets around very active stars (undergoing frequent flares) are probably never at a steady-state but are constantly and permanently altered by flare events.
These important variations are detectable on the synthetic spectra as deviations from the steady-state can be up to 1200 ppm (Fig.\ref{spectrum_recu}).

\section{Progress in terms of modelling}\label{sec:prog_mod}

Most chemical models of hot/warm exoplanets have involved one-dimensional vertical profile models with a specified input temperature profile. However, recent model developments have extended this to take into account 1) consistency of the chemistry with the temperature profile and 2) extend the model dimension to include the effects of horizontal transport.

\subsection{Coupling chemistry-thermal profile}

As was shown previously (Sect. \ref{section:model_parameters}), various model parameters can have a large impact on the chemical abundances. In particular, for a given elemental composition, the temperature profile largely determines calculated chemical abundances through the temperature-dependence of the rate constants. In addition, if vertical mixing is taken into consideration, the location of the quench point is a play-off between the chemical timescale and mixing timescale, which depend on the temperature and mixing strength (K$_{zz}$) respectively.

\begin{figure}[h!]
\begin{center}
\includegraphics[width=0.9\columnwidth]{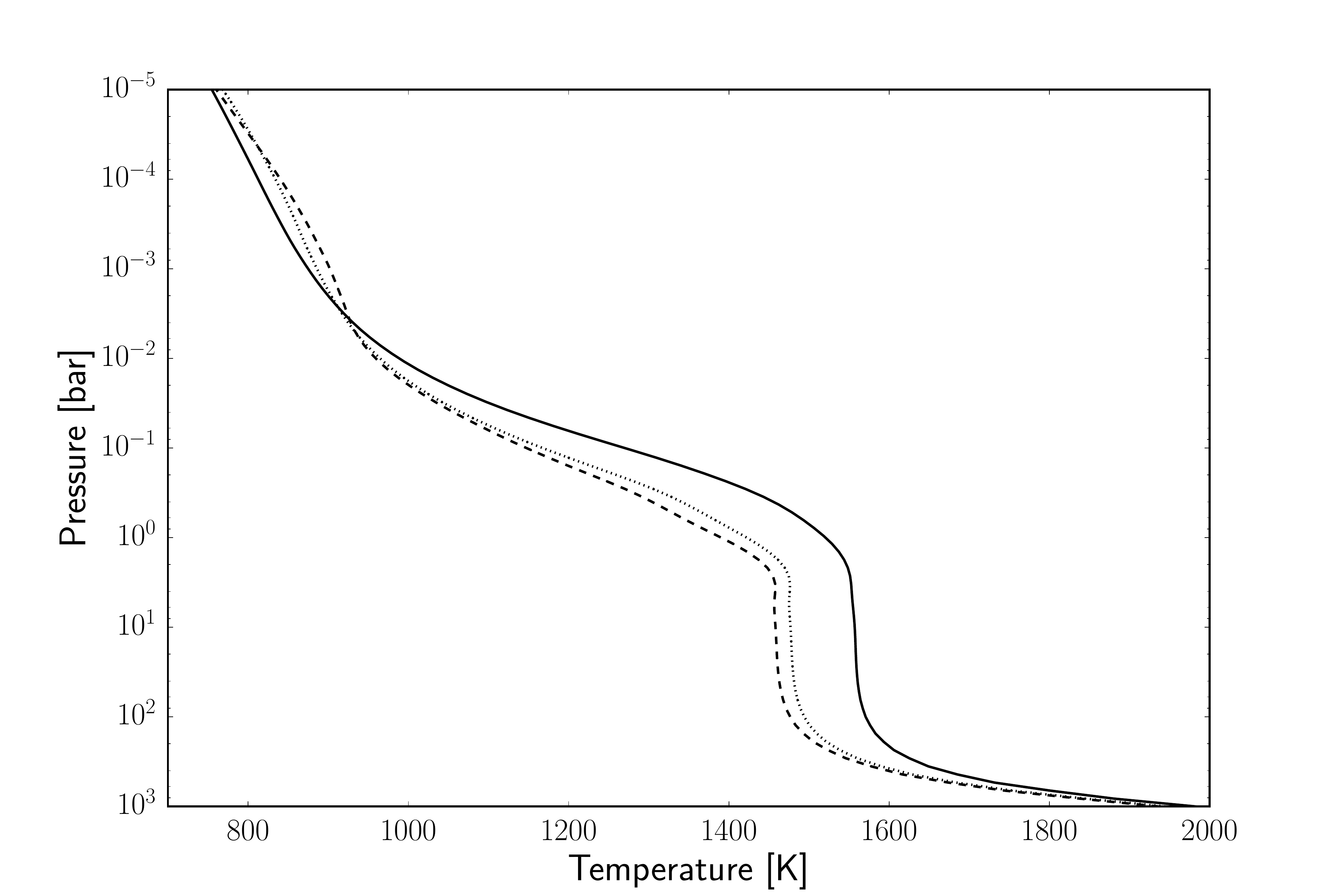}
\caption{{\label{figure:pt_hd189}
The pressure-temperature profiles derived for HD 189733b assuming chemical equilibrium (dashed), and with vertical mixing and photochemistry included for $K_{zz}$ = 10$^9$cm$^2$s$^{-1}$ (dotted) and $K_{zz}$ = 10$^{11}$cm$^2$s$^{-1}$. From \cite{drummond2016}, reproduced with permission \copyright ESO. %
}}
\end{center}
\end{figure}

However, the temperature profile is in turn dependent on the chemical composition, as it controls the opacity and hence the absorption and emission of radiation. \cite{drummond2016} used a 1D radiative-convective equilibrium model, which includes a chemical kinetics scheme, using the chemical network of \cite{Venot_2012}, to solve for the temperature profile which is consistent with non-equilibrium abundances (i.e. including vertical mixing and photochemistry). 
It was found that the process of vertical mixing can have a strong influence on the temperature profile compared with the temperature profile consistent with chemical equilibrium, depending on the strength of the mixing. Figure \ref{figure:pt_hd189} shows the pressure-temperature profiles for a model of HD 189733b, with profiles consistent with both chemical equilibrium and non-equilibrium (vertical mixing and photochemistry). Depending on the strength of the mixing, non-equilibrium chemistry can have an important influence on the temperature profile; locally increasing the temperature by up to 100 K. 

This change in the temperature due to non-equilibrium abundances has a feed back on the abundances themselves, through the temperature dependent rate constants. In this case, this leads to larger CH$_4$ abundances and smaller CO abundances, compared with the model where the temperature profile is held fixed. The impact of photochemistry on the temperature profile is very small, since photochemistry is only important at low pressures ($P<10^{-5}$ bar) where the atmosphere is optically thin.

\begin{figure}[h!]
\begin{center}
\includegraphics[width=0.48\columnwidth]{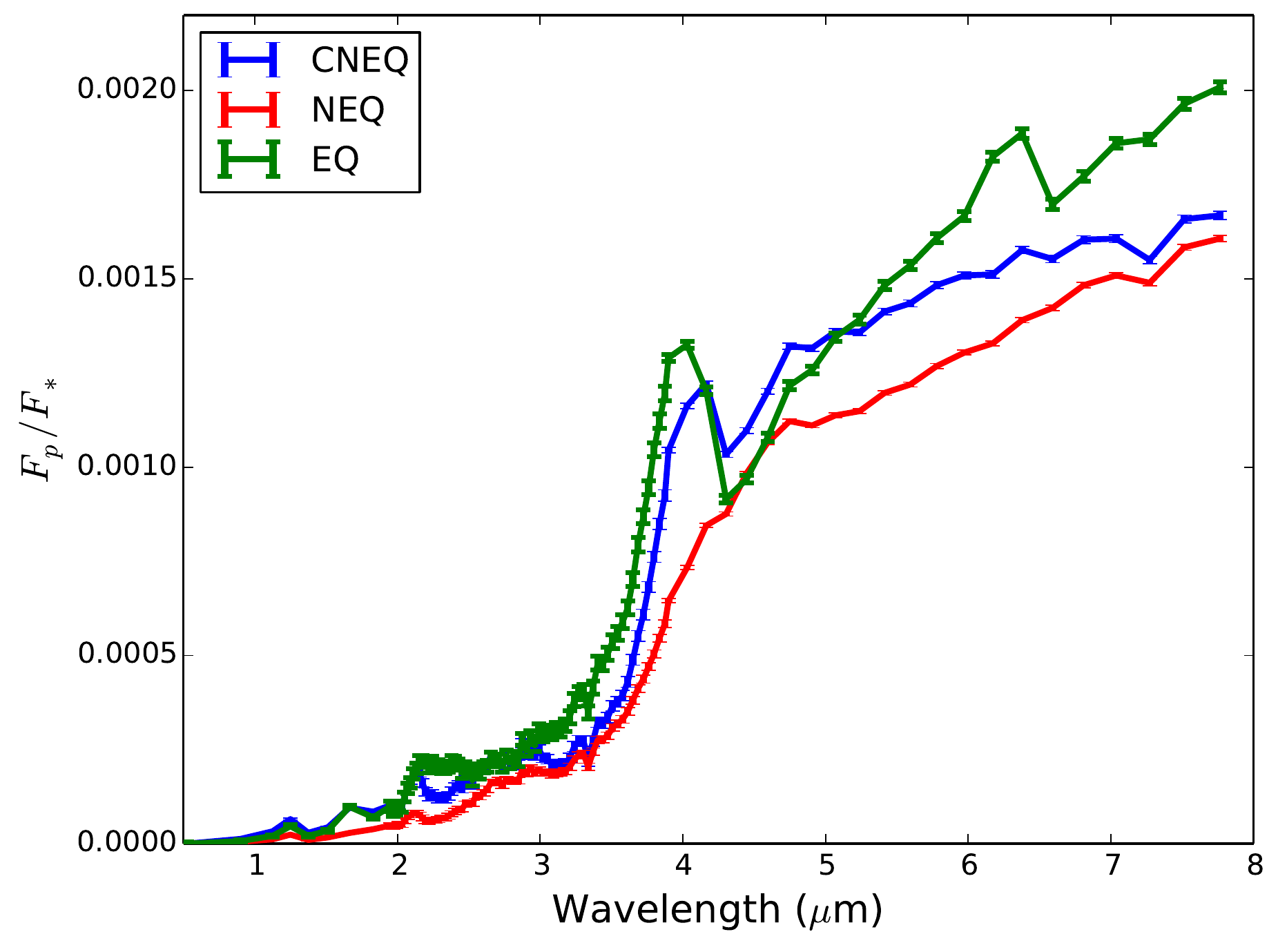}
\includegraphics[width=0.48\columnwidth]{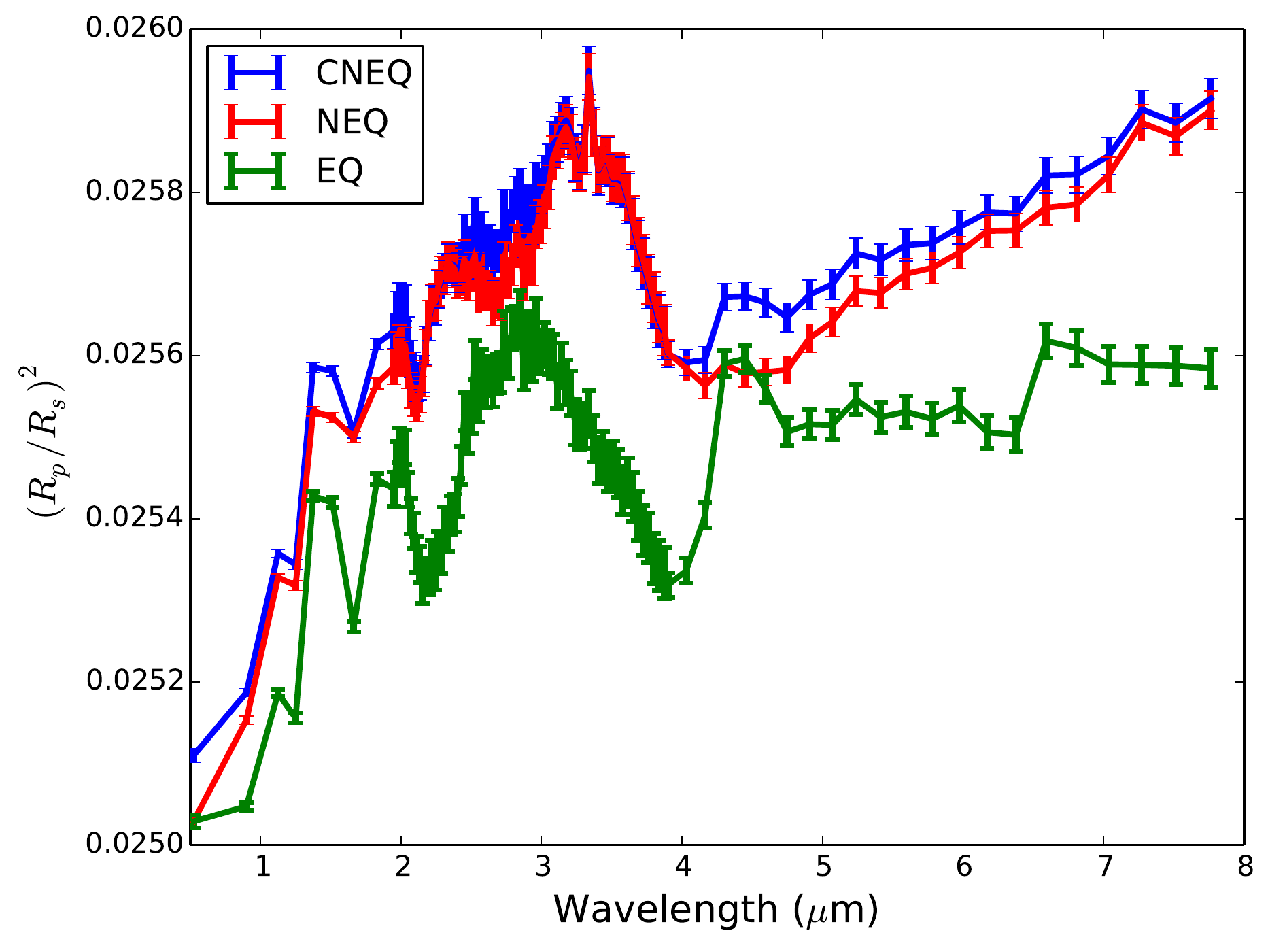}
\caption{{\label{figure:spectrum_hd189_ARIEL}
The simulated emission (\textit{left}) and transmission (\textit{right}) spectra of HD 189733b that will be observed by ARIEL, for $K_{zz}$ = 10$^{11}$cm$^2$s$^{-1}$, showing the model assuming chemical equilibrium (EQ, green), the model with non-consistent non-equilibrium chemistry (NEQ, red) and the model with consistent non-equilibrium chemistry (CNEQ, blue). Spectra are binned to ARIEL resolution, as explained in Table \ref{tab:char_ARIEL}. The error bars correspond to a SNR of 21.%
}}
\end{center}
\end{figure}

Calculating non-equilibrium abundances consistently with the temperature structure also has an impact on the simulated observations. Figure \ref{figure:spectrum_hd189_ARIEL} shows the emission and transmission spectra for the same model, comparing chemical equilibrium, non-consistent non-equilibrium and consistent non-equilibrium cases. The spectra have been binned to the resolution of ARIEL (see Table \ref{tab:char_ARIEL}). Note these spectra have been recalculated (see Section \ref{sec:rt}) from the thermal and composition profiles presented in \citet{drummond2016}.

The emission spectra of the three cases (equilibrium, consistent non-equilibrium and non-consistent non-equilibrium) are all different to each other. However, Fig. \ref{figure:spectrum_hd189_ARIEL} shows that of the two non-equilibrium cases, the consistent model (blue line) is closer to the equilibrium model (green line) for most wavelength regions; there are however, exceptions to this, for example around 4.5 $\mu$m. This means that, depending on the wavelength, previous studies using a non-consistent approach could have overestimated the effect of non-equilibrium chemistry on the emission spectrum. In transmission the consistent and non-consistent non-equilibrium cases give broadly similar results.

The difference between the consistent and non-consistent models can be understood in terms of energy balance of the atmosphere. The photosphere is the location in the atmosphere where the optical depth is approximately unity, and is where energy can freely escape to the top of the atmosphere. The pressure level of the photosphere is dependent on wavelength due to the wavelength-dependent opacity which, in turn, is determined by the chemical composition. The energy flux from the atmosphere , at a particular wavelength, is determined by the atmospheric temperature at the photospheric pressure level. 

The main consequence of non-equilibrium chemistry, in this particular case, is to increase the abundances of CH$_4$ and NH$_3$ via vertical quenching. The corresponding increase in opacity shifts the photosphere to lower pressures, and lower temperatures, for the wavelengths at which these species absorb, changing the energy flux and the shape of the emission spectrum. While the spectral energy flux from the atmosphere may change, due to a changing composition, the total energy flux (i.e. integrated over wavelength) must balance the total energy input to the model atmosphere. Here, irradiation and internal heating are the only sources of energy to the atmosphere and are constant for all three cases.

\citet{drummond2016} found that the consistent non-equilibrium model gives the same integrated emitted energy flux as the equilibrium model, to within the expected accuracy of the calculations, as expected while the non-consistent non-equilibrium model shows a 39$\%$ decrease compared with the equilibrium model. The non-consistent model does maintain energy balance. To conserve energy balance the consistent calculation allows the temperature profile (in this case heats up) to evolve mitigate the effect of a shifting photosphere. We refer the reader to \citet{drummond2016} for a more detailed discussion.

The variations between these three models will be differentiable with ARIEL. With one single observation (SNR=7), the resolution of ARIEL is sufficient to separate the chemical equilibrium model from the two others, both in emission and transmission. Concerning the non-equilibrium chemistry models, separating the consistent one from the non-consistent one requires different SNR depending on the type of observation: for an observation during the secondary eclipse, a SNR of 7 allows to distinguish them for $\lambda>$ 4 $\mu$m. In the range 2--4 $\mu$m, about nine observations are required, resulting in a SNR of 21. On primary transit observations, the two models are more difficult to disentangle. For $\lambda <$ 2 $\mu$m and between 4 and 5 $\mu$m, the two spectra can be separated with a SNR of 14, but elsewhere a SNR $>$ 21 is required. Note, however, that in the range [2 -- 2.5] $\mu$m and [3 -- 4] $\mu$m, the two non-equilibrium chemistry models give very similar spectra that can't be distinguished, even with very high signal-to-noise ratios. Overall, this study shows that considering consistency between chemistry and temperature is very important in atmospheric modeling, as the variations on spectra between the different models are greater than the uncertainties on the future observations of ARIEL.

It is possible that studies not taking into account consistency of chemistry and temperature may have overestimated the impact of non-equilibrium chemistry on the spectrum, due to the models not conserving energy. An additional process not considered in \cite{drummond2016} is adjustments to the thermal profile due to energy changes associated with chemical reactions, which may also play a role in linking the chemical composition with the thermodynamics of the atmosphere.

\subsection{Horizontal mixing}

All of the results presented so far have used a 1D code that approximates the atmosphere as a single column. However, short-period exoplanets are expected to be tidally locked, with large day-night temperature contrasts and fast horizontal winds. The effects of this three dimensional temperature structure and dynamics cannot be consistently captured with a 1D model. Coupling a complex chemical model to a 3D circulation model is a large computational challenge.

To date, two attempts have been made to include the effects of horizontal mixing in hot Jupiter atmospheres: 1) coupling a complex 3D model with a very simplified chemical scheme \citep{CS2006} and 2) coupling a complex chemical scheme with a simplified circulation model, a pseudo-2D model \citep{agu2012,agu2014}. In this latter case, the 1D atmospheric column rotates as a solid body to mimic a uniform zonal wind; i.e. time is taken as an {\it effective} second spatial dimension.

\begin{figure}[h!]
\begin{center}
\includegraphics[width=1\columnwidth]{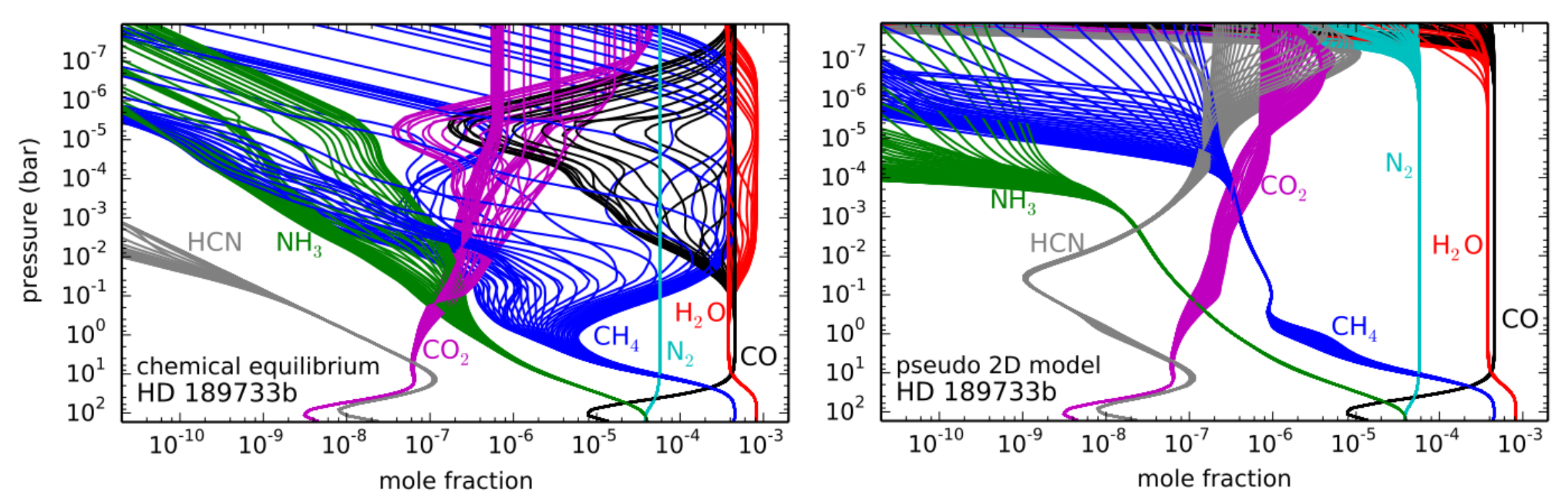}
\caption{{\label{pseudo2D}
Vertical abundances profiles for the atmosphere of the hot Jupiter HD 189733b. The composition corresponding to the thermochemical equilibrium is on the left. The composition calculated with the pseudo 2D model is on the right. For each species, the different lines correspond to different longitudes spanning the 0--360\degree~range. Adapted from \citet{agu2014}, reproduced with permission \copyright ESO.%
}}
\end{center}
\end{figure}

In their model, \citet{CS2006} include only H$_2$O, CO, and CH$_4$, which are the main carbon and oxygen bearing species. Their goal is to investigate the CO/CH$_4$ interconversion in the atmosphere of HD 209458b. They argue in terms of timescales estimates and find that vertical mixing is more important than horizontal mixing. They find that vertical quenching from a deep layer leads to a horizontally uniform composition for lower pressures. On the other hand, \citet{agu2014} study the detailed chemical composition of HD 209458b and HD 189733b, with a chemical scheme involving more than 100 species. In both planets, they find that horizontal mixing is more important, and that the chemistry is quenched horizontally, with the chemistry of nightside being `contaminated' with that of the dayside (see Fig. \ref{pseudo2D}). Understanding the differences between the results of these two studies may require extending capabilities to consistently include a full chemical kinetics scheme in a 3D circulation model: a difficult, but important, challenge. 

Planetary spectra are usually computed with 1D radiative-transfer codes. Because of the nature of these models, the longitudinal/latitudinal variations of the temperature and the chemical composition are neglected. \citet{agu2014} investigate in which extent these longitudinal variations can affect the emission and transmission spectra. Because the chemical composition is horizontally quite homogenized, they find that the variability of composition with phase does not affect the emission spectrum, neither the transmission spectrum. The differences between the ingress and egress transmission spectra on one side, and the emission spectra at different phases on the other side are in reality due to the temperature variations (see Figs. 18 and 19 in \citealt{agu2014}). The only change due to the chemical composition gradient is observed around 4.3 $\mu$m and is due to the variation of CO$_2$ abundance.

\begin{figure}[h!]
\begin{center}
\includegraphics[width=0.49\columnwidth]{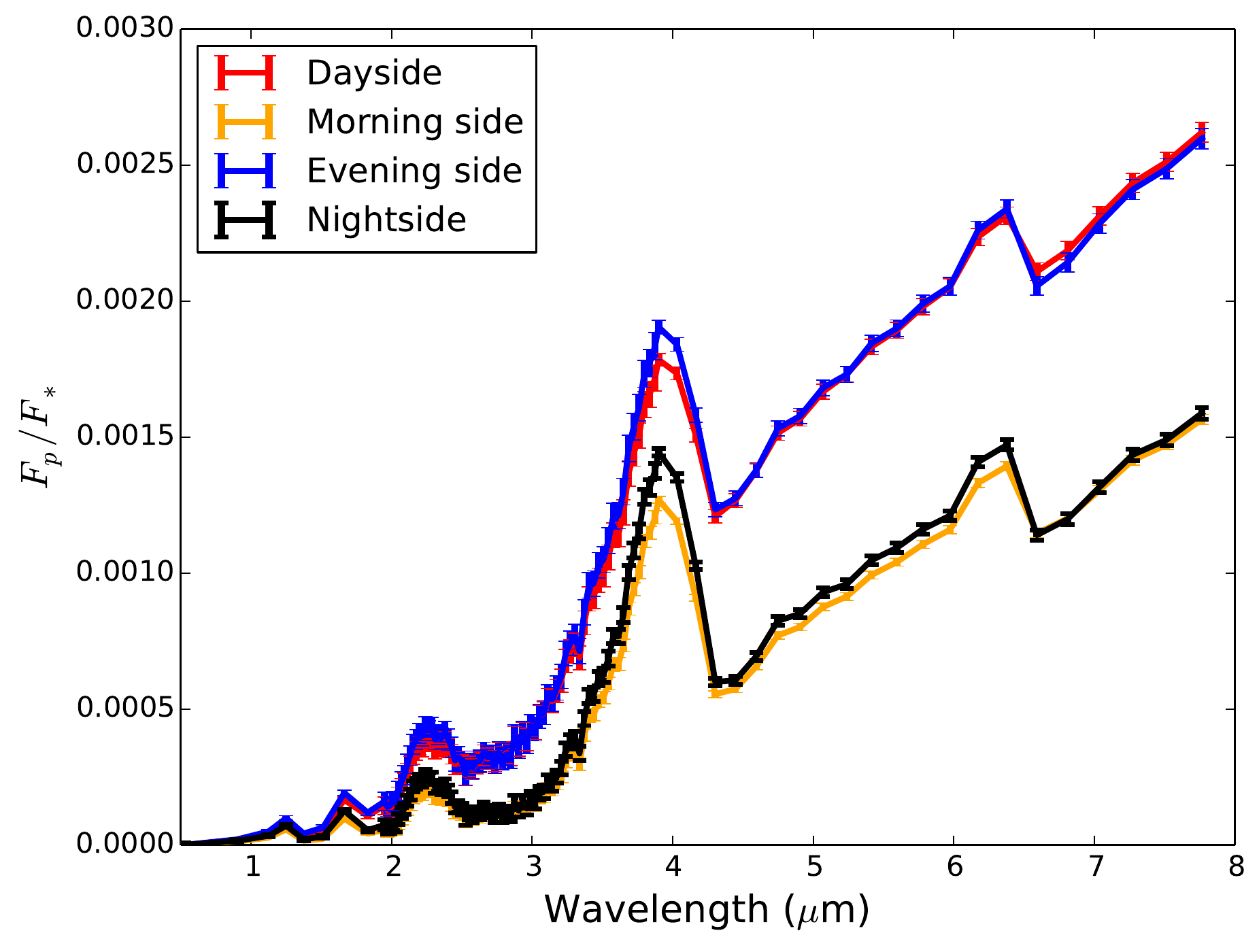}
\includegraphics[width=0.49\columnwidth]{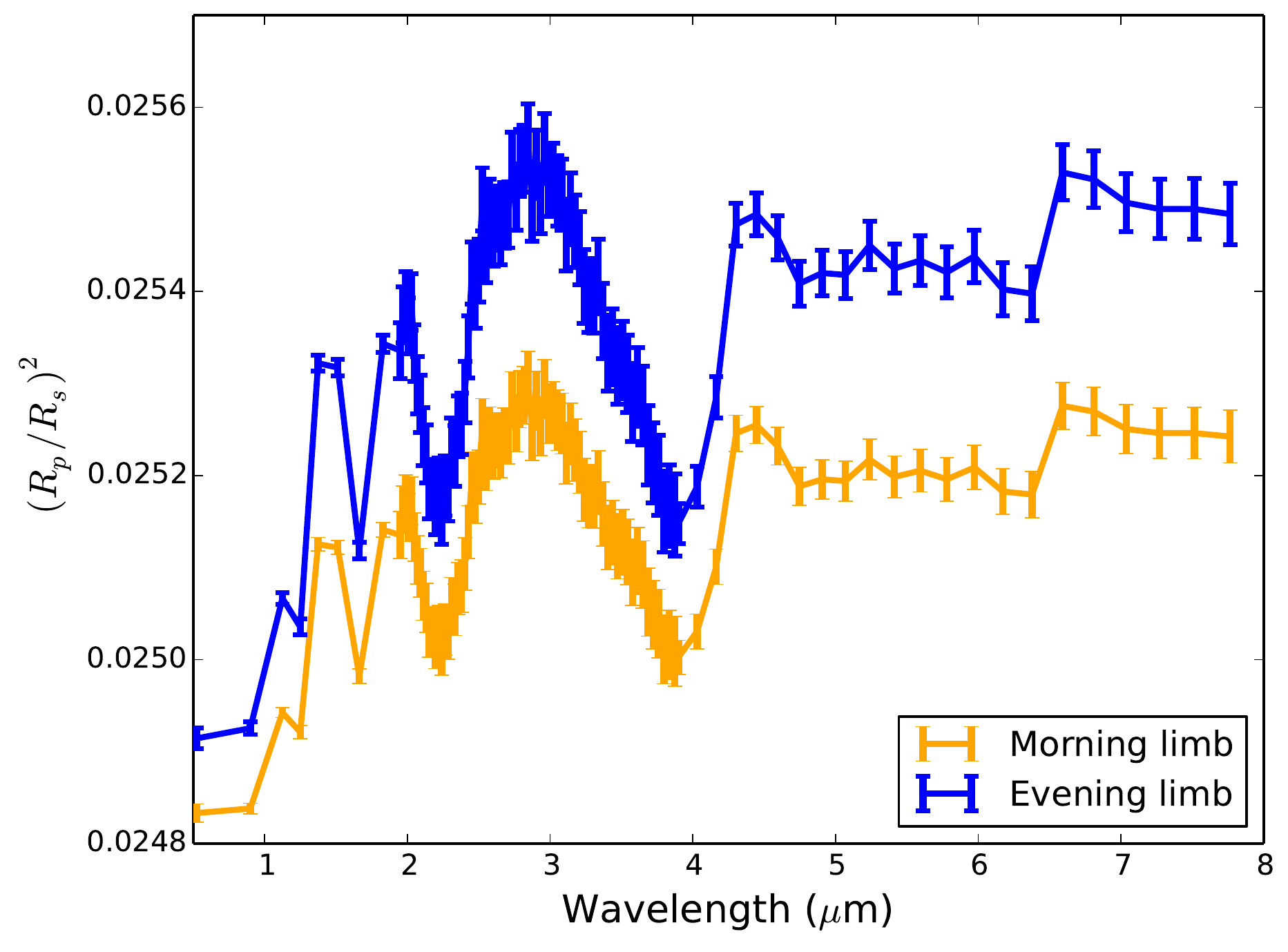}
\caption{{\label{spectra_pseudo2D}
Synthetic emission (\textit{left}) and transmission (\textit{right}) spectra for HD 189733b, with the chemical compositions calculated by the pseudo 2D model. They are binned to ARIEL resolutions, as explained in Table \ref{tab:char_ARIEL}. The error bars correspond to a SNR of 14.
}}
\end{center}
\end{figure}

To estimate if the changes of the planetary spectra with phase can be detected with ARIEL we calculated the emission spectra corresponding to the morning, day, evening, and night sides, as well as the transmission spectra corresponding to the morning and evening limbs (Fig. \ref{spectra_pseudo2D}).
We binned them to the resolution of ARIEL following the characteristics of Table \ref{tab:char_ARIEL}. For secondary eclipse, one can observe two groups of emission spectra: the first one is made of the morning and night sides, and the second one gathers the day and evening sides. These two groups can be easily differentiated with one single observation, which corresponds to a SNR of 7 in the case of HD 189733b (see Table \ref{tab:SNR}). However, it is more difficult to disentangle the two components of each groups because they are very close. The feature around 4 $\mu$m can be identified with a SNR of 14 (i.e. four observations), but elsewhere it remains difficult to separate the spectra of each group.\\
For primary eclipse observations, we found that the transmission spectra of the morning and evening limbs are differentiable on the overall spectral range from a SNR of 7 (i.e. one single observation). Regarding these results, it is certain that ARIEL will allow us to better understand hot Jupiters, as it will have the technical capacity to determine horizontal and vertical variations of the thermal structure of their atmospheres.

\section{Need for experimental data at high temperature}

Whatever the degree of sophistication of the atmospheric kinetics model (one dimension with constant thermal profile or chemistry-consistent thermal profile, two dimensions, and probably in the future three dimensions), of fundamental importance is to use data corresponding to the properties of this system. A model will always provide a result, and conclusions are then very easy to draw. However, if the input data used in the model are wrong, conclusions will be false, however advanced the model is. 

With respect to the chemistry calculations, the term "input data" includes all the data necessary to calculate kinetics and photodissociations: reaction rates, absorption cross-sections, and quantum yields. Concerning the reaction rates, it has already been presented in Section \ref{sec:1Dmodel} that, taking advantage of decades of intensive work in the field of combustion, \citet{Venot_2012} and \citet{Venot_2015} developed chemical schemes adapted to high temperature and validated experimentally.

But concerning absorption cross-sections and quantum yields, no such work exists already. Currently, all chemical models dealing with high temperatures don't have another choice but to use data at ambient temperature (or at 350-400 K in the best cases). Thus, an important uncertainty exists in the modelling of photochemistry.

In this context, an ambitious project has been developed at the Laboratoire Interuniversitaire des Syst\`emes Atmosph\'eriques (France). It consists of measuring the absorption cross-sections of the most important species of planetary atmospheres at temperatures relevant for exoplanets. Thanks to an experimental setup that enables the gas to be heated to high temperatures (currently up to 1000 K), measurements are being performed in synchrotron facilities (BESSY, Germany or SOLEIL, France).
The absorption cross-section of carbon dioxide, $\sigma_{CO2}$, has been studied by \citet{Venot_2013} and \citet{Venot_2018} in the range [115-230] nm, up to 800 K.
\begin{figure}[h!]
\begin{center}
\includegraphics[width=0.9\columnwidth]{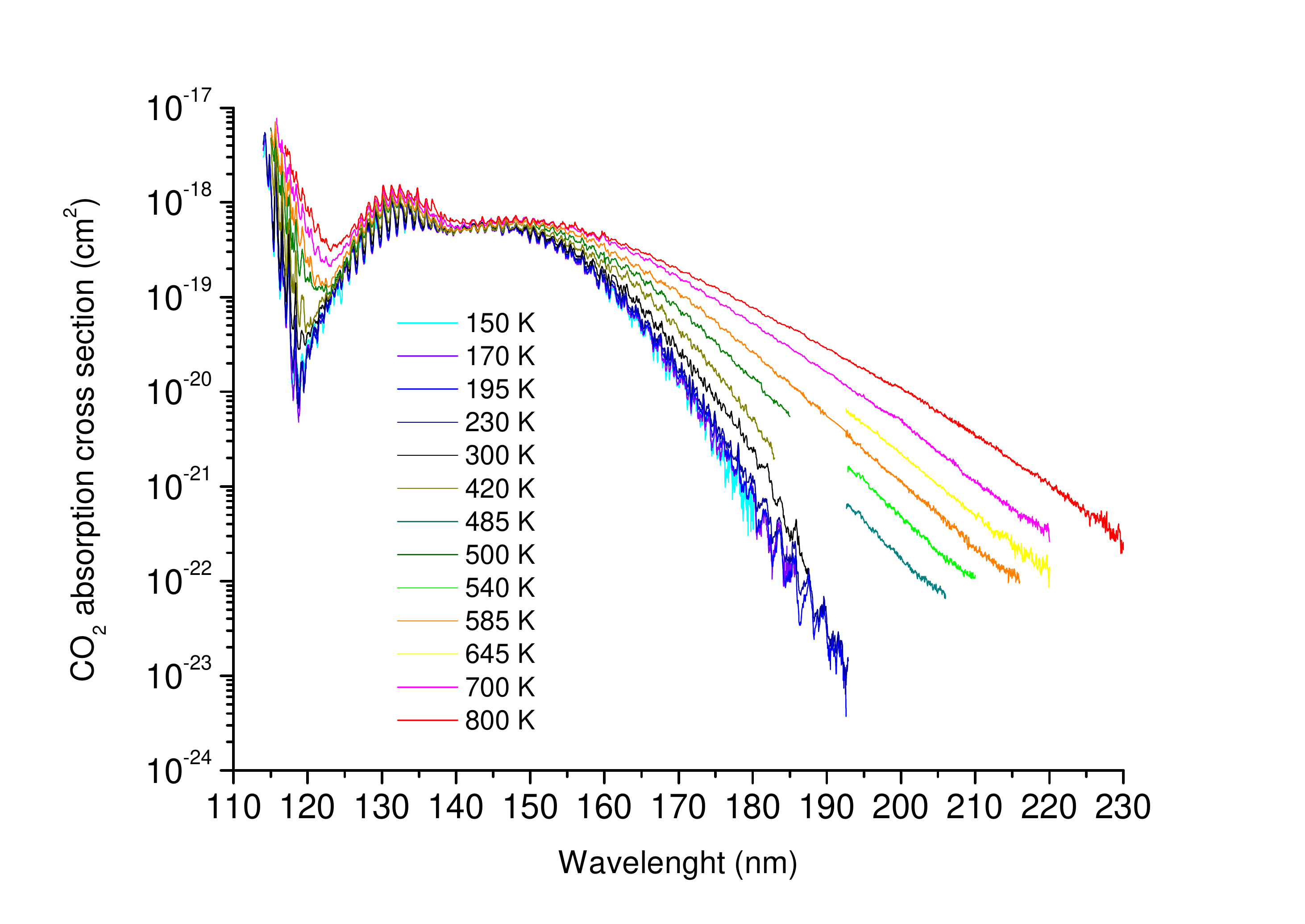}
\caption{{\label{CO2_sigma}
VUV absorption cross sections of carbon dioxide at different temperatures from 150 to 800 K. Adapted from \citet{Venot_2013, Venot_2018}, reproduced with permission \copyright ESO.}}
\end{center}
\end{figure}
The increase of the absorption together with the temperature is spectacular (see Fig. \ref{CO2_sigma}). Moreover, the use of this more accurate data in atmospheric models significantly affects the predicted chemical composition (see Fig. \ref{fm_venot2018}). The abundance of CO$_2$ is modified mainly in two regions: around 10$^{-3}$ mbar and around 10 mbar. Between these two regions, the destruction of CO$_2$ by photolysis is compensated by the production of CO$_2$ through the reaction CO + OH $\longrightarrow$ CO$_2$ + H.
\begin{figure}[h!]
\begin{center}
\includegraphics[width=0.49\columnwidth]{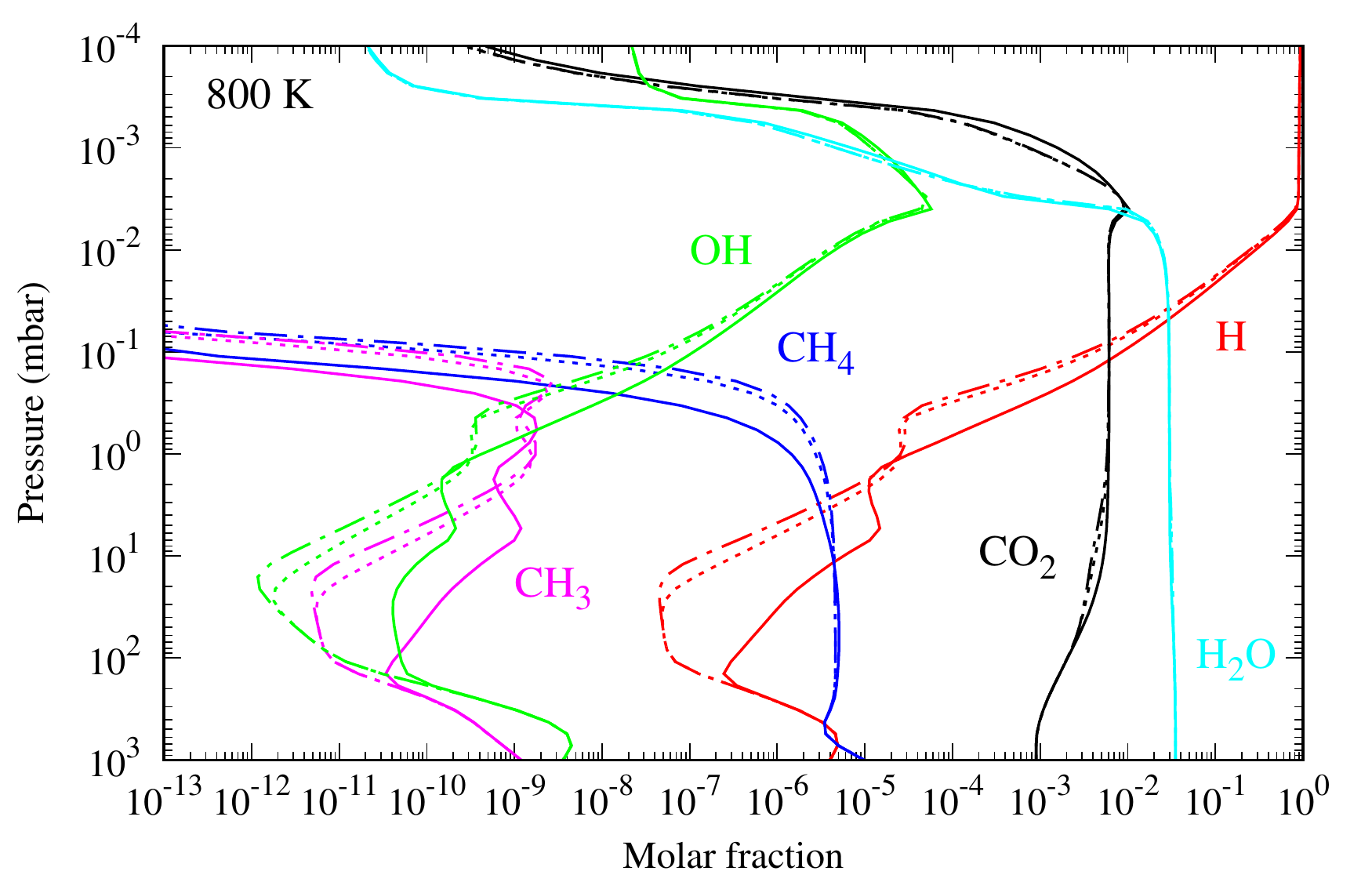}
\includegraphics[width=0.49\columnwidth]{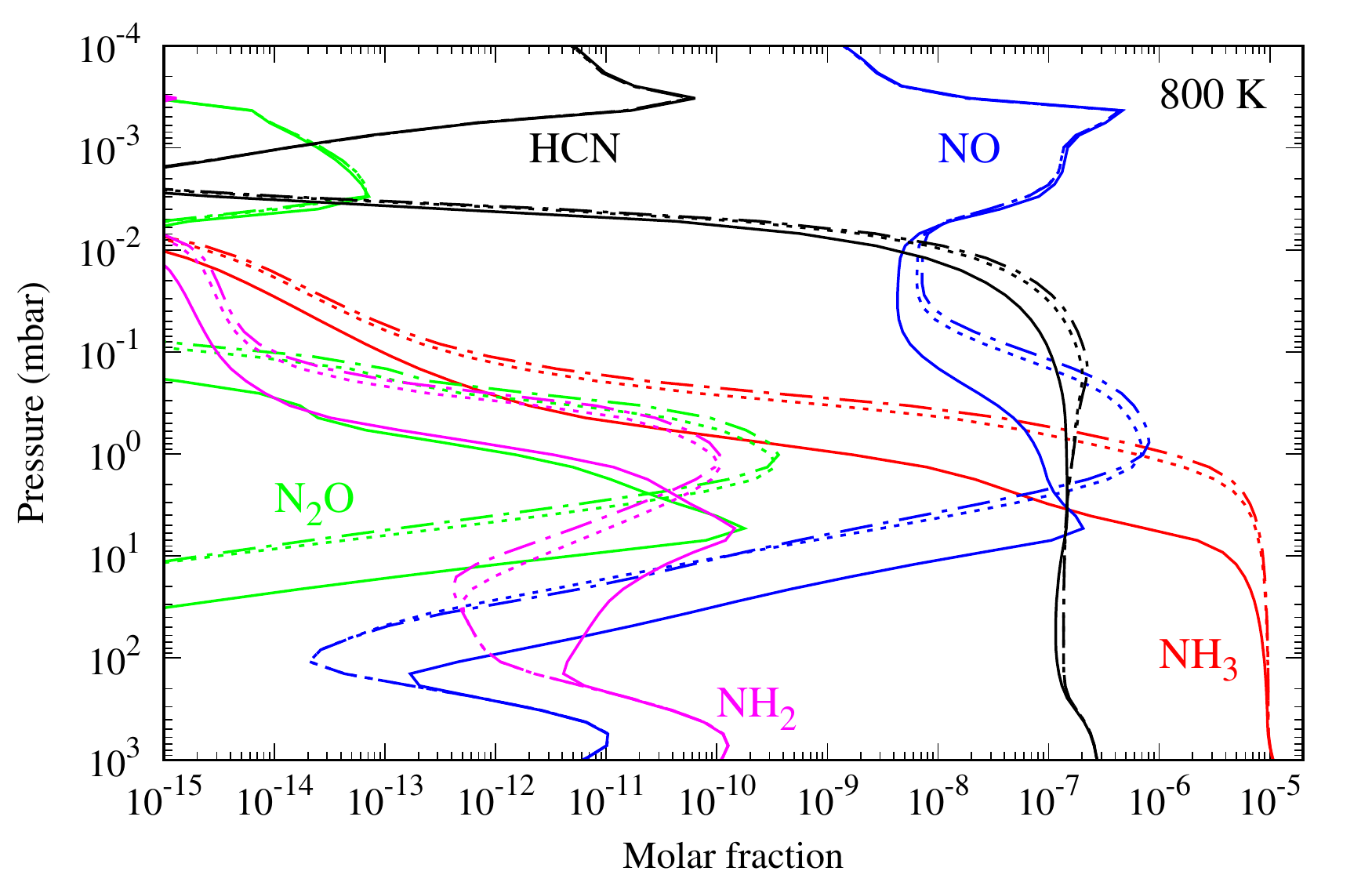}
\caption{{\label{fm_venot2018}
Vertical abundances of several species when using the absorption cross section of CO$_2$ at 300 K (full line) or at 800 K (dotted line). The atmospheric temperature between 10$^{-4}$ and 10 mbar is 800 K. The dot-dashed lines correspond to a model using an analytical formula for the absorption of CO$_2$ determined by \citet{Venot_2018}. From \citet{Venot_2018}, reproduced with permission \copyright ESO.}}
\end{center}
\end{figure}
Changing only the absorption cross section of CO$_2$, many species see their abundances varying by several orders of magnitude (NH$_3$, H, CH$_4$,\dots). This confirms the urgency of acquiring a complete database of cross-sections at high temperature.

\begin{figure}[h!]
\begin{center}
\includegraphics[width=0.49\columnwidth]{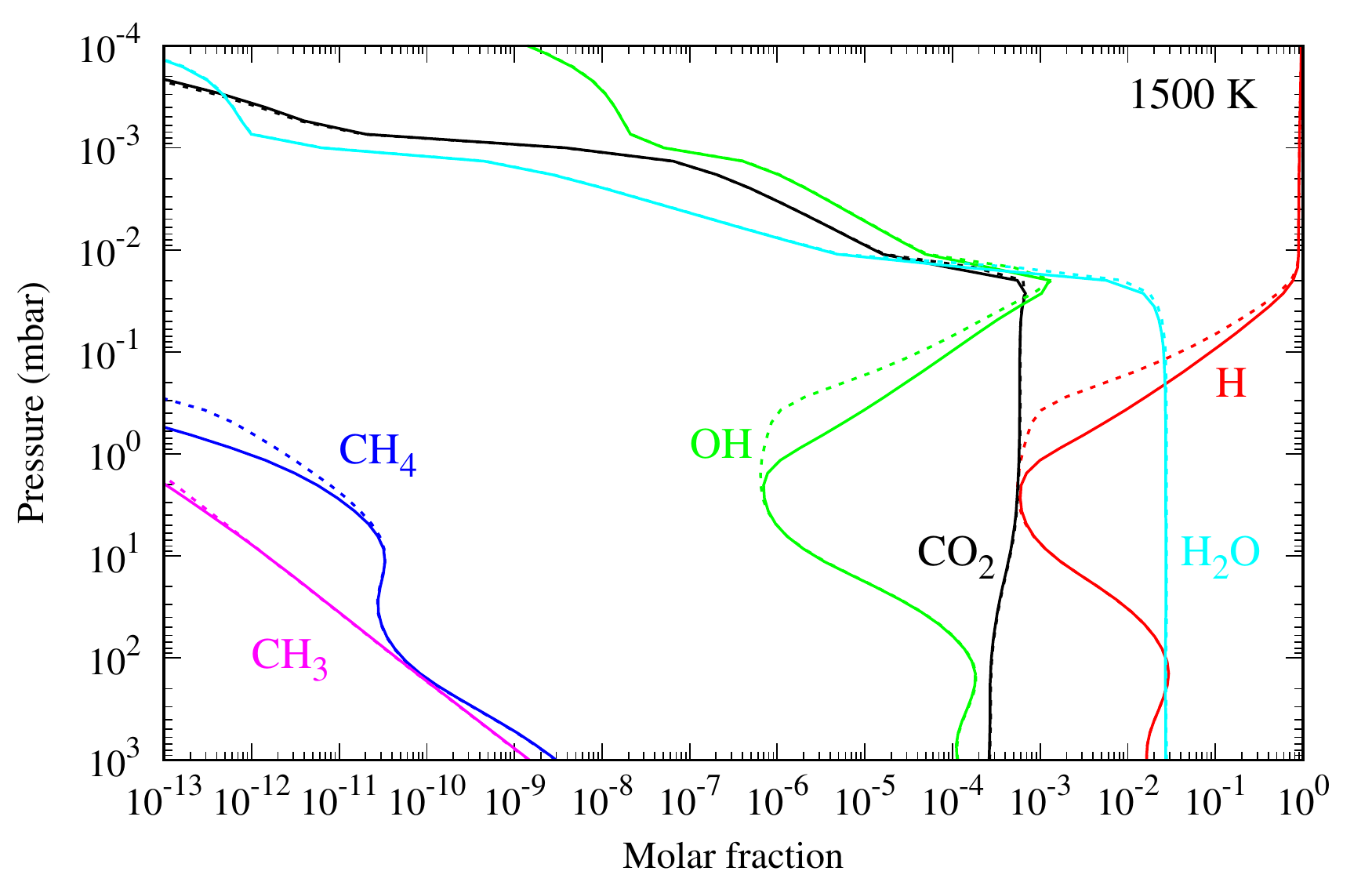}
\includegraphics[width=0.49\columnwidth]{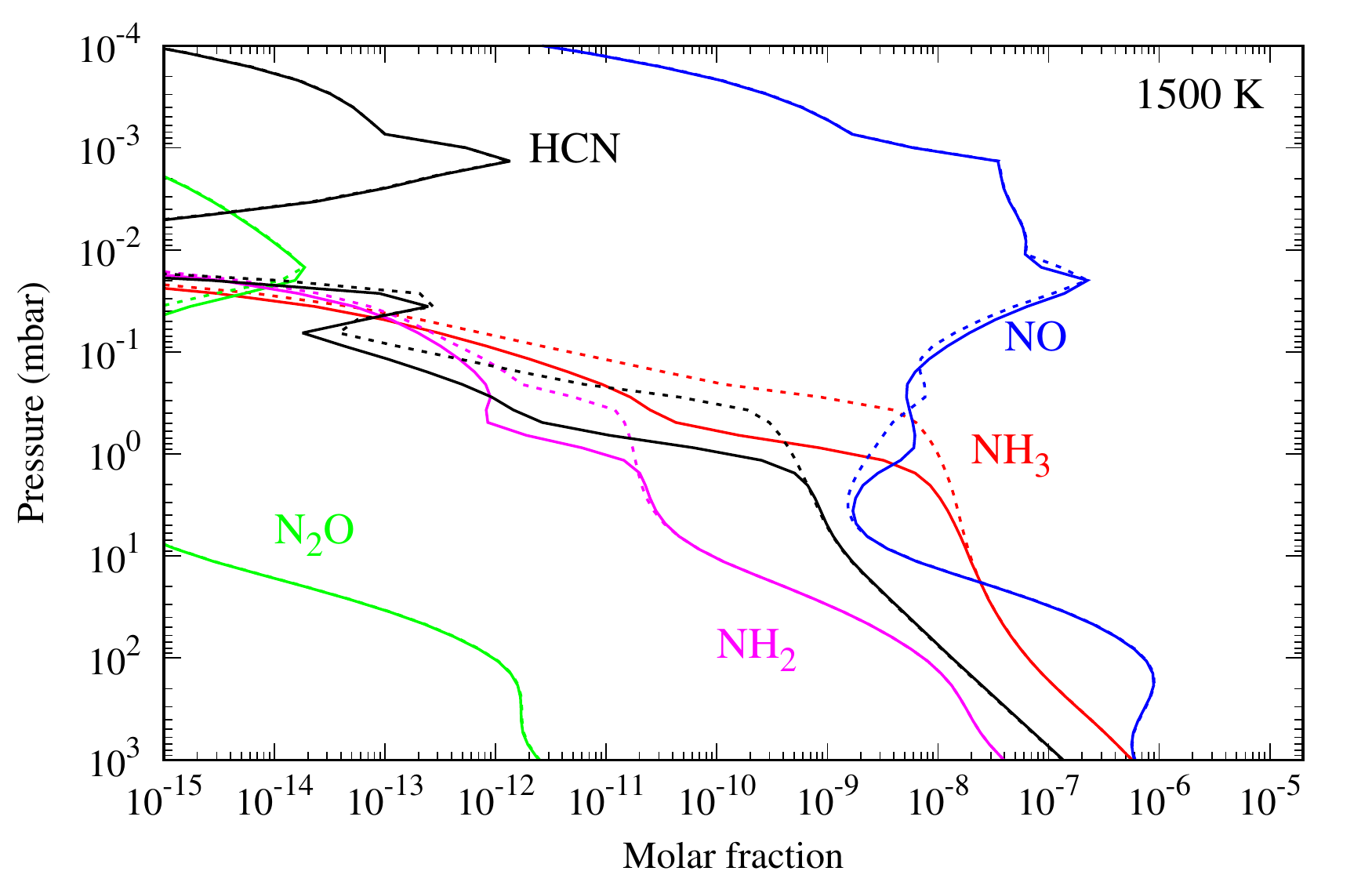}
\caption{{\label{fm_venot2018_1500}
Vertical abundances of several species when using the absorption cross section of CO$_2$ at 300 K (full line) or at 1500 K (dotted line). The absorption at 1500 K is calculated with an analytical formula. The atmospheric temperature between 10$^{-4}$ and 10 mbar is 800 K. From \citet{Venot_2018}, reproduced with permission \copyright ESO.}}
\end{center}
\end{figure}

\begin{figure}[h!]
\begin{center}
\includegraphics[width=\columnwidth]{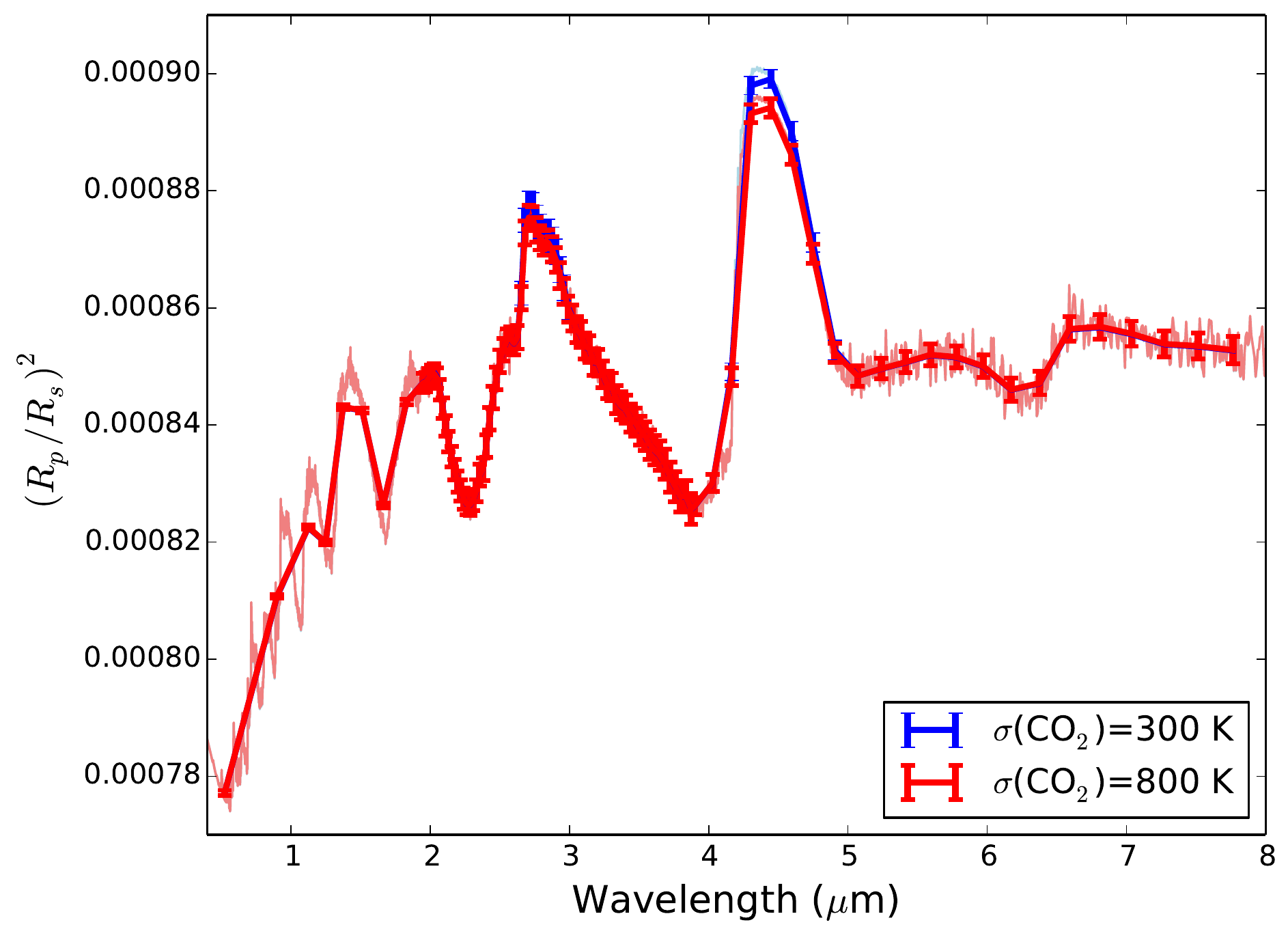}
\caption{{\label{ARIEL_bin_venot2018}
Synthetic transmission spectra of the atmosphere at 800 K corresponding to the models using the absorption cross section of CO$_2$ at 300 K (blue) or at 800 K (red). The bold spectra are binned to ARIEL resolution, as explained in Table \ref{tab:char_ARIEL}. The error bars correspond to a SNR of 20. The fainter spectra have a higher resolution (R=300), constant in wavelength.}}
\end{center}
\end{figure}

In \cite{Venot_2018}, the authors also determine a parametrization of the absorption cross section of CO$_2$ that permits to calculate the absorption of this molecule (more exactly the continuum of the absorption) at any temperature in the wavelength range [115-230] nm. Thus, they estimate the absorption of CO$_2$ at 1500~K and use this data to study a very hot atmosphere (i.e. T = 1500 K between 10$^{-4}$ and 10 mbar). They find that in this very hot atmosphere, the contribution of photodissociations to the chemical composition is smaller than in cooler atmospheres. Consequently, even if the absorption of CO$_2$ is much more important at 1500~K than at 800~K, using $\sigma_{CO2}$(1500 K) instead of $\sigma_{CO2}$(300 K) generates less variations than in the atmosphere at 800 K (see Fig.~\ref{fm_venot2018_1500}).
Variations of the chemical composition induced by the change of VUV absorption cross section in the photochemical model have noticeable consequences on the transmission spectrum in the case of a 800K atmosphere only. High-resolution transmission spectra are shown in \cite{Venot_2018}. Here we recalculated these spectra and binned them to the resolution of ARIEL assuming a signal-to-noise of 20. To estimate the noise, we took the characteristics of the star HD 128167 (F star, T$_{eff}$ = 6600 K, d = 15.83 pc, mag$_k$=6) since its stellar flux has been used in the photochemical model. \cite{Venot_2018} showed that the spectra corresponding to the models using the CO$_2$ absorption cross section at 300 K or at 800 K present differences at three locations: at 2.7, 4.3, and 14.9 $\mu$m. These differences are quite small, but the departure at 4.3 is distinguishable on ARIEL observations if they are performed with a SNR $\ge$ 20 (see Fig.\ref{ARIEL_bin_venot2018}). The range around 14.9 $\mu$m will not be observed by ARIEL.

In view of these results, it appears more urgent to study other molecules at temperatures up to $\sim$1000 K rather than performing measurements at higher temperatures.
Measurements of other species are in progress, such as NH$_3$, C$_2$H$_2$, and HCN, and should be released soon.

\section{Chemical modeling and ARIEL}

Whilst significant developments are still required, the current state-of-the-art models provide us with an understanding of how various parameters and processes (temperature, metallicity, mixing, etc.) can affect the atmospheric chemical compositions of hot exoplanets. Indeed, for a given atmosphere it is possible to determine whether the atmosphere is likely to be described by chemical equilibrium or whether non-equilibrium processes will have a significant effect. If non-equilibrium effects are likely, then the use of kinetics models can predict the abundances of the chemical species and investigate the effect of these non-equilibrium abundances on the observable spectra. We have shown in this manuscript that the high precision of ARIEL will be able to discriminate relatively easily (with one or a few tens observations) the different assumptions that can be done of atmospheres (metallicity, C/O ratio, temperature, etc.).

The limiting factor to a wider and deeper knowledge of exoplanet atmospheres is currently the limited precision, resolution and amount of observations. The coming of next-generation instruments will provide much higher precision spectra, which will constrain the models to a higher degree. However, the number of observations which can be made with a finite amount of telescope time is strictly limited. Therefore, ARIEL is eagerly awaited as it is the only project which can deliver a large survey of atmospheric spectra. Such a database of observed spectra for a large number of planets with different (and similar) properties is crucial to understanding how the planet properties, stellar type and orbital configurations alter the atmospheric chemistry; and of course, this understanding can be only be achieved by combining theory and observations. 

By studying the chemical composition of the atmospheres of exoplanets, we can hope to gain some insight into the bulk compositions of such planets and eventually to their formation and evolution histories. The huge diversity of exoplanets discovered to date indicates that such a problem can only be tackled by considering a large sample of planets, which ARIEL will provide. We will finally be able to understand why so many exoplanetary systems currently detected appear radically different to our own Solar system, and indeed, to determine whether the Earth and the Solar system are truly unique.

\begin{acknowledgements}
All figures extracted from previous publications have been reproduced with permission. The authors deeply thank the anonymous referee for his/her comments that greatly improve the manuscript. O. V. thanks the CNRS/INSU Programme National de Plan\'etologie (PNP) for funding support. B.D. acknowledges funding from the European Research Council (ERC) under the European Unions Seventh Framework Programme (FP7/2007-2013) / ERC grant agreement no. 336792. I.P.W. acknowledges funding from the European Research Council (ERC) under the European Unions Seventh Framework Programme (FP7/2007-2013) / ERC grant agreement no. 758892. T. Z. is supported by the European Research Council (ERC) project ExoLights (617119) and from INAF trough the "Progetti Premiali" funding scheme of the Italian Ministry of Education, University, and Research. Y.M. greatly appreciates the CNES post-doctoral fellowship program and support for travel funding.
\end{acknowledgements}


\bibliographystyle{spbasic}
\bibliography{Exp.Astro_ARIEL}

\end{document}